\documentclass{aa}
\usepackage{txfonts}
\usepackage{natbib}
\bibpunct{(}{)}{;}{a}{}{,}
\usepackage{graphicx}
\newcommand{\teff}{$T_{\rm{eff}}$}
\newcommand{\logg}{$\log g$}
\newcommand{\lL}{\ifmmode \log \frac{L}{L_{\sun}} \else $\log \frac{L}{L_{\sun}}$\fi}

\newcommand{\kms}{km~s$^{-1}$}
\newcommand{\msun}{M$_{\sun}$}

\begin{document}

\title{On the maximum helium content of multiple populations in the globular cluster NGC~6752}
\author{F. Martins\inst{1}
\and W. Chantereau\inst{2}
\and C. Charbonnel\inst{3,4}
}
\institute{LUPM, Universit\'e de Montpellier, CNRS, Place Eug\`ene Bataillon, F-34095 Montpellier, France  \\
\and
Université de Strasbourg, CNRS, UMR7550, Observatoire astronomique de Strasbourg, 67000 Strasbourg, France \\
\and
Department of Astronomy, University of Geneva, Chemin Pegasi 51, 1290 Versoix, Switzerland
\and IRAP, CNRS UMR 5277 \& Université de Toulouse, 14 avenue Edouard Belin, 31400 Toulouse, France
}

\offprints{Fabrice Martins\\ \email{fabrice.martins@umontpellier.fr}}

\date{Received / Accepted}

\abstract
{Multiple populations in globular clusters are usually explained by the formation of stars out of material with a chemical composition that is polluted to different degrees by the ejecta of short-lived, massive stars. But the nature of the ``polluters'' remains elusive. Different types of stars have been proposed to account for the observed chemical patterns of multiple populations. Among other things, these differ by the amount of helium they spread in the surrounding medium.}
{In this study we investigate whether the present-day photometric method used to infer the helium content of multiple populations indeed gives the true value or underestimates it by missing very He-rich, but rare stars. This check is important to discriminate between the different polluter scenarios. We focus on the specific case of NGC~6752.}
{We compute atmosphere models and synthetic spectra along isochrones produced for this cluster for a very broad range of He abundances covering the predictions of the different scenarios, including the extreme case of the fast-rotating massive star (FRMS) scenario. We use the same abundances in isochrones and atmosphere models to ensure consistency. We calculate synthetic photometry in HST filters best suited to study the helium content. We subsequently build synthetic clusters with various distributions of stars. We finally determine the maximum helium mass fraction of these synthetic clusters using a method similar to that applied to observational data. In particular, we select nonpolluted and very He-rich stars from the so-called chromosome map.}
  {We re-determine the maximum helium mass fraction Y in NGC~6752, and find a value consistent with published results. We build toy models of clusters with various distributions of multiple populations and ensure that we are able to recover the input maximum Y. We then build synthetic clusters with the populations predicted by the FRMS scenario and find that while we slightly underestimate the maximum Y value, we are still able to detect stars much more He-rich than the current observed maximum Y. This result still holds even in synthetic clusters that contain less He-rich stars than predicted by the FRMS scenario. It is easier to determine the maximum Y on main sequence stars than on red giant branch stars, but qualitatively the results are unaffected by the sample choice.}
{We show that in NGC~6752 it is unlikely that stars more He-rich than the current observational limit of about 0.3 (in mass fraction) are present. }
{}


\authorrunning{Martins et al.}
\titlerunning{Synthetic photometry of NGC~6752}

\maketitle

%%%%%%%%%%%%%%%%%%%%%%%%%%%%%%%%%%%%%%%%%%%%%%%%%%%%%%%%%%%%%%%%%%%%%%%%%%%%%%%%%%%%%%%%%%%%%%%%%%%%%%%%%%%%%%%%%%%%%%%%%%%%%%%
%%%%%%%%%%%%%%%%%%%%%%%%%%%%%%%%%%%%%%%%%%%%%%%%%%%%%%%%%%%%%%%%%%%%%%%%%%%%%%%%%%%%%%%%%%%%%%%%%%%%%%%%%%%%%%%%%%%%%%%%%%%%%%%
\section{Introduction}
\label{s_intro}

Globular clusters (GCs) host multiple populations of stars characterized by different spectroscopic and photometric signatures. Anti-correlations between the abundances of several elements such as C-N, Na-O, and sometimes Mg-Al have been reported at the surface of various types of stars, from the main sequence (MS) to the red giant branch (RGB) and the asymptotic giant branch (AGB) \citep[e.g.,][]{Cohen1978,Peterson1980,sneden92,Kraft1994,Thevenin2001,gratton07,Gratton2019,Meszaros2015,carretta15,Carretta2019,Johnson2016,Pancino2017,Wang2017,Masseron2019}. These differences in chemical composition are thought to explain the various sequences observed in color-magnitude diagrams (CMDs) built with specific filters sensitive to atomic and molecular lines of the elements listed above \citep[e.g.,][]{bedin04,piotto07,Bowman2017,nardiello18,marino19}. The discovery of these peculiarities (variations in chemical composition and discrete sequences in CMDs) turned GCs from textbook examples of clusters born in a single star forming event into complex structures the origin of which remains elusive.

The standard explanation\footnote{Alternative possibilities exist but all rely at least partly on nucleosynthesis products of AGB or massive stars, see \citet{marcolini09,elmegreen17}.} to these puzzling properties is the pollution of the original proto-cluster gas by a population of rapidly evolving stars more massive than the present-day GC members. Indeed, the abundance patterns observed are typical outcomes of nucleosynthesis at about 75 MK. More precisely, hydrogen burning through the CNO cycle, together with the Ne-Na and the Mg-Al chains, fully explains the anti-correlations \citep{1988ATsir1525...11K,Denisenkov1990,prantzos07,Prantzos2017}. Potential polluters experiencing such nucleosynthesis phases are $\sim$5-7.5~\msun\ AGB stars \citep[e.g.,][]{ventura01,Ventura2009}, massive stars either rotating fast \citep{decressin07,DecressinCM2007}, in binary systems \citep{demink09,Izzard2013}, or in a red supergiant phase \citep{szecsi19}, and super-massive stars \citep{denis14}. Whatever the nature of the polluter, it should be objects present at early times of the GC history that have spread their nucleosynthesis products in the surrounding gas and out of which the stellar populations we see today were formed. Depending on the degree of mixing of the nucleosynthesis products with pristine gas, a range of chemical composition is predicted for the newly formed stars. The polluters have long disappeared because of their relatively high mass, and only low-mass stars either from the initial population formed out of original proto-GC gas or from the second population formed out of polluted material are still present. We usually refer to these two populations of stars as 1P (or 1G) and 2P (or 2G), standing for first and second population (generation), respectively.

One of the key differences in the predictions of the scenarios that invoke the different polluters listed above is the amount of helium that is inevitably ejected together with other nucleosynthesis products. Indeed, as hot hydrogen-burning is needed to explain the observed abundance patterns, helium is naturally produced and released too. On one hand, the fast-rotating massive star (FRMS) models of \citet{decressin07}, developed  specifically to reproduce the observations of the GC NGC~6752, predict a wide range of Y for the ejecta, up to Y=0.8, and the test case\footnote{The study focuses on one binary system made of a 20~\msun\ primary and a 15~\msun\ secondary star orbiting each other on a 12-day orbit, and with individual rotation periods synchronized on the orbital period.} of binary evolution presented by \citet{demink09} also predicts a wide range of Y, up to 0.63.
On the other hand, models of \citet{doherty14} indicate that nucleosynthesis in massive AGB stars should produce material with helium mass fraction of about 0.35--0.40, as a result of second dredge-up.  On the other hand, supermassive stars
can potentially release material rich in hot hydrogen-burning products but only mildly enriched in helium \citep{denis14}, with a helium mass fraction close to that of the proto-cluster gas in the case where they form through the so-called runaway collision scenario \citep{Gieles2018}.

Measuring the helium surface abundance from spectroscopy is not possible in most GC stars. Only hot horizontal branch (HB) stars display weak \ion{He}{i} lines, and these are difficult to model and interpret. However, the helium surface abundance in these stars, with effective temperatures higher than $\sim$11 000~K, is affected by atomic diffusion  \citep[e.g.,][]{Michaud08,Quievy09}. Thus, the measured helium no longer represents the original chemical composition of the stars. The vast majority of GC stars are too cool to display any helium line in their spectra. Nevertheless, helium enrichment is detected in a few hot HB stars with $\Delta$Y ---the difference between the highest Y of 2P stars and Y of 1P stars--- generally not larger than $\sim$0.1 \citep{villanova12,marino14}, although \citet{pasquini11} and \citet{dupree13} report a Y difference of up to 0.17 between two HB stars, of NGC~2808 and $\omega$~Cen, respectively.

A change in helium mass fraction affects the internal structure as well as the color and the brightness of stars because the opacity is modified. As a consequence, stars with the same mass, age, and metallicity but different Y have different effective temperatures (\teff; the higher Y, the higher \teff; e.g., \citealt{IbenFaulkner1968,chantereau15,CassisiSalaris2020}). Their spectral energy distributions (SEDs) therefore peak at different wavelengths: stars with high Y have more flux at shorter wavelengths, and thus bluer colors. Consequently they are located to the left of stars with smaller Y in CMDs (e.g., \citealt{sbordone11} and Sect.~\ref{s_sed}). Variations in the helium content naturally lead to a widening of classical branches (MS, RGB, HB, AGB) in CMDs \citep[e.g.,][]{Rood1973,Norris1981,DAntonaCaloi2004,DAntona2005ApJ,chantereau16}. In addition, the associated variations in (among others) C, N, and O abundances impact specific colors built with photometric filters encompassing lines sensitive to these elements. This further increases the separation between stars in CMDs, leading to the observed multiple and discrete sequences \citep[e.g.,][]{Lardo2012,marino19}. 

Comparison of observed and theoretical colors in filters mostly sensitive to \teff, and thus Y, has been used to   indirectly constrain the amount of helium present at the surface of GC stars \citep[e.g.,][]{piotto07,milone15}. \citet{king12} report a maximum helium mass fraction of about 0.39$\pm$0.02 in $\omega$~Cen based on the analysis of MS stars. Using the Hubble Space Telescope (HST) ultra-violet (UV) survey of GCs \citep[HUGS][]{piotto15,nardiello18}, \citet{milone18} determined Y in 57 GCs. $\Delta$Y ranges from nearly 0 to 0.124, corresponding to a maximum Y of about 0.38. \citet{milone18} also show that $\Delta$Y correlates with the cluster mass, with higher Y being determined in more massive clusters.

Another indirect constraint on the helium content of GCs comes from the morphology of their HB. Its different shape in clusters with similar general properties, such as M3 and M13 \citep[e.g.,][]{ferraro97}, is difficult to explain. A spread in Y of HB stars is a viable possibility \citep{rood73} and quantitative determinations give $\Delta$Y between 0.02 and 0.15 on the HB, depending on the cluster \citep{Caloi2005,lee05,dicri10,valcarce16,tailo16,denis17,vandenberg18,chantereau19}. 

The presence of hot HB stars in GCs of early-type galaxies is also thought to be responsible for the existence of a so-called UV upturn \citep{gr90}. This feature refers to the increase of flux below $\sim$2500 \AA\ in galaxies that no longer form stars and that are made of low-mass MS and post-MS stars. These objects have SEDs that rapidly drop short of about 3500~\AA. Hot HB stars, such as those seen in Galactic GCs, may explain the UV fluxes. \citet{ali18} argued that if such hot HB stars are present because of a high helium content (as suggested by \citealt{Meynet2008} for elliptical galaxies), there should be a redshift above which they would no longer be observed, having not yet sufficiently evolved. The position of this transition redshift can be used to constrain Y in hot HB stars. \citet{ali18} showed that a value of $\sim$0.45 would be compatible with the observed disappearance of the UV upturn with redshift.

Thus, modern estimates of Y in GCs indicate that the very helium-rich stars predicted in particular by the FRMS scenario are not detected. However, a key specificity of the FRMS model highlighted by \citet{chantereau15} is that such He-rich objects evolve faster and differently compared to more classical stars. In particular, \citet{chantereau16} showed that in a GC that would have been formed under the FRMS scenario, the distribution of stars quickly falls as Y increases (at ages typical of GCs, i.e., 9 to 13.5 Gyr). In practice, there should be little to no He-rich stars on the RGB and AGB of the cluster NGC~6752, for which the models of Chantereau et al. were tailored. This also naturally explains the lack of Na-rich AGB stars in some GCs \citep{Campbell2013,Wang2016,Wang2017}, although observations reveal that the presence of 2P AGB stars can be affected by more than one  factor \citep{Wang2017}.

The question therefore naturally arises as to whether the absence of very He-rich stars (i.e., Y$>$0.4) is a robust observational fact, or these stars simply escaped detection due to their exceptionally small number. In view of the discriminating nature of the helium content for various scenarios, it is important to investigate this potential issue, which we aim to do in the present study. In practice, we build on the work of \citet{chantereau16} to compute synthetic clusters with the predicted distribution of multiple populations for NGC~6752. We compute synthetic spectra consistently with isochrones, produce synthetic photometry and CMDs, and perform determinations of the maximum Y values of these synthetic clusters in order to see if we recover or miss the highly enriched populations predicted by the model of Chantereau et al.

Our paper is organized as follows: Sect.~\ref{s_specphotom} presents the computation of synthetic spectra and photometry. We then discuss the behavior of our models with special emphasis on the effects of surface abundances on colors. We subsequently compare our predicted photometry to observations in various CMDs before moving to the determination of the maximum helium content in NGC~6752 and synthetic clusters in Sect.~\ref{s_maxHe}. We discuss our results in Sect.~\ref{s_disc} and give our conclusions in Sect.~\ref{s_conc}.

%%%%%%%%%%%%%%%%%%%%%%%%%%%%%%%%%%%%%%%%%%%%%%%%%%%%%%%%%%%%%%%%%%%%%%%%%%%%%%%%%%%%%%%%%%%%%%%%%%%%%%%%%%%%%%%%%%%%%%%%%%%%%%%
\section{Spectral synthesis and synthetic photometry}
\label{s_specphotom}

In this section we first present our computations of synthetic spectra and the associated photometry (Sect.~\ref{s_meth}). We then describe the effects of stellar parameters and surface abundances on both the SEDs and synthetic colors (Sect.~\ref{s_sed}). We build synthetic CMDs and compare them with observations (Sect.~\ref{s_compobs}).

%%--------------------------------------------------------------------------
\subsection{Method}
\label{s_meth}

\begin{figure}[t]
\centering
\includegraphics[width=9cm]{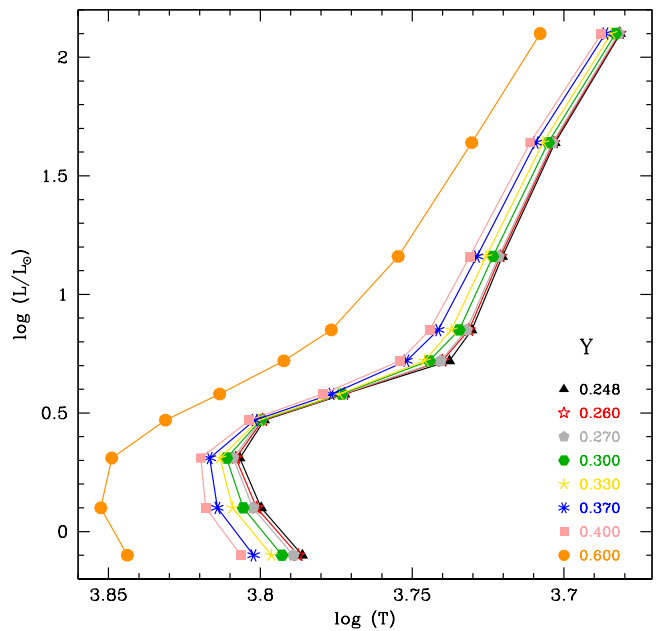}
\caption{Hertzsprung-Russell diagram with the isochrones of \citet{chantereau15}. Different symbols and colors stand for different initial composition, especially different helium mass fraction. Symbols correspond to the points for which atmosphere models have been calculated.}
\label{hr_theo}
\end{figure}

We proceeded as in \citet{martins20} to compute synthetic photometry. We relied on the evolutionary tracks and isochrones computed by \citet{chantereau15} with the code STAREVOL.
These models are based on the FRMS scenario \citep{decressin07} for the formation of multiple stellar populations in GCs. Each isochrone is characterized by a set of abundances that directly comes from the FRMS predictions. We refer to \citet{chantereau15} for details. Each set of abundances is tailored to reproducing 1P and various 2P stars in NGC~6752, in which the abundances vary according to different degrees of pollution by FRMS. Figure~\ref{hr_theo} shows the Hertzsprung-Russell diagram (HRD) with the isochrones we consider in the present work. Each isochrone is labeled according to its helium mass fraction (Y), but we stress that abundances of other elements are also varied (in particular, nitrogen is increased and carbon and oxygen are depleted when Y increases; see Appendix~\ref{ap_chem}). We stress that the isochrones have been recomputed for a metallicity of $[Fe/H]=-1.53$, which is slightly higher than the value of -1.75 used in \citet{chantereau15}. This change was made to better match the metallicity of NGC~6752. 

Along each isochrone we computed atmosphere models and synthetic spectra at ten points (see Fig.~\ref{hr_theo}) using the codes ATLAS12 \citep{kur14} and SYNTHE \citep{kur05}. Our computations include the so-called predicted lines, that is, lines for which at least one of the energy levels comes from quantum mechanics calculation and not from laboratory measurements. These lines thus have approximate wavelengths, but their opacities have been shown to be of prime importance to reproducing observed SEDs \citep{coelho14}. We adopted a microturbulent velocity of 1~\kms\ in all our calculations. 

To convert the HRD into CMDs we computed synthetic photometry in the Vegamag system: 

\begin{equation}
  mX = -2.5 log (F_X/F^{Vega}_X) = -2.5 log (F_X) + ZP^{Vega}_X
\label{eq_mX}
,\end{equation}

\noindent where $mX$ is the magnitude in the X filter and ZP the zero point. The average flux $F_X$ over the passband X was calculated according to \citet{bohlin14}:

\begin{equation}
F_X = \frac{\int \lambda F_{\lambda} R_X d\lambda}{\int \lambda R_X d\lambda}
,\end{equation}

\noindent where $R_X$ is the transmission curve of filter X\footnote{Transmission curves were retrieved from the Spanish VO \url{http://svo2.cab.inta-csic.es/svo/theory/fps3/}.}.
The zero point $ZP^{Vega}_X$ in Eq.~\ref{eq_mX} was calculated using the Vega STScI reference spectrum\footnote{File alf\_lyr\_stis\_010.fits from \url{ftp://ftp.stsci.edu/cdbs/current_calspec/}.} and the appropriate transmission curve.

Finally, for comparison with observations, we assumed a distance modulus of 13.18 for NGC~6752 based on the \emph{Gaia} DR2 determination \citep{helmi18}. This value is consistent with those found by \citet{harris96} (13.19), \citet{renzini96} (13.05) , and \citet{gratton03} (13.24).
Prior to synthetic photometry calculations, we reddened our SEDs using a color excess $E(B-V) = 0.060$ and A$_V$=3.2, adopting the extinction law of \citet{ccm89}.
This choice best reproduces the turn-off (TO) and sub-giant region in the  m814W (m606W-m814W) CMD (see for instance Fig.~\ref{cmd}). This is slightly larger than the value of 0.046$\pm$0.005 reported by \citet{gratton05} but close to the value of \citet{shleg98}:  0.056.

%%--------------------------------------------------------------------------
\subsection{Spectral energy distributions}
\label{s_sed}

In this section we describe the effects of chemical composition on the SED. The goal is to identify spectral regions that are affected by certain species. We refer to \citet{sbordone11} or \citet{milone20}, among other recent studies, for similar descriptions.

\begin{figure}[h]
\centering
\includegraphics[width=9cm]{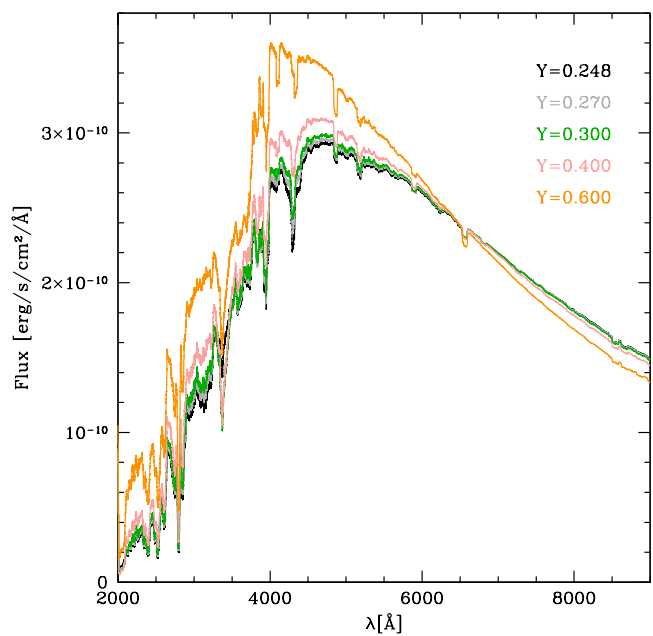}
\caption{Spectral energy distributions of some of the models with \lL~=~0.85 but different chemical compositions (and thus different \teff\ and \logg). Different colors correspond to different models, identified by their helium mass fraction.}
\label{comp_sed_L0p85}
\end{figure}

Figure~\ref{comp_sed_L0p85} shows a selection of models at the bottom of the RGB, with luminosities equal to $10^{0.85}$~L$_{\odot}$ (see Fig.~\ref{hr_theo}). These spectra have different \teff, surface gravities and chemical compositions (see Table~\ref{tab_chem}). Because of the higher \teff\ in higher Y models, the SED peak is shifted towards shorter wavelengths. In the optical region, the slope of the SED becomes steeper (faster decline with wavelength) which translates into bluer colors. Variations in individual abundances of light elements (CNO) also changes the strength of absorption lines, mostly below 4500~\AA. 

\begin{figure}[t]
\centering
\includegraphics[width=9cm]{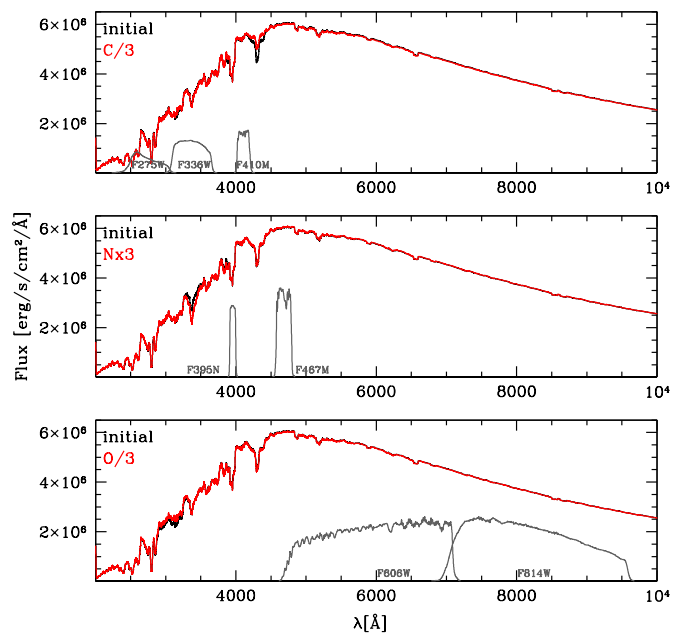}
\caption{Spectral energy distribution of two models with similar parameters except the C, N, and O abundances: the red line shows a model with C/H reduced by a factor three (top panel), N/H increased by a factor three (middle), and O/H reduced by a factor three (bottom) compared to the initial model in black. The other parameters are: \teff\ = 5375 K, \logg\ = 3.37, Y=0.248. The gray solid lines indicate the transmission curves of the HST WFC3/UVIS F275W, F336W, F395N, F410M, F467M, and ACS/WFC F606W and F814W.}
\label{comp_CNO}
\end{figure}

To better separate the effects of light elements and helium on the SED, we show in Fig.~\ref{comp_CNO} examples of spectra in which the abundances of C, N, and O have been varied by a factor of three, all other parameters being kept constant. A reduction of the carbon abundance translates into a weaker CH absorption between 4200 and 4400 \AA. Conversely, an increase in the nitrogen content leads to a stronger NH absorption at 3300-3500 \AA. The CN band around 3900 \AA\ is less affected. The OH absorption between 2800 and 3300 \AA\ reacts to a change in the oxygen content. Consequently, photometry in the HST filters F275W and F336W is affected by C, N, and O abundances \citep{sbordone11,milone18,CassisiSalaris2020}. These filters have been used to build the so-called super color C$_{410}$=(m275W-m336W)-(m336W-m410M) \citep{milone13}. This photometric diagnostic has been shown to be a powerful tool for separating multiple populations (see also following section). Filters F395N, F467M, F606W, F814W and to some extent F410M are relatively insensitive to variations in C, N, and O abundances. An important result of Fig.~\ref{comp_CNO} is that C, N, and O abundances do not affect the global shape of the SED, but only specific wavelength regions that include molecular bands \citep{milone13,Dotter2015}.

\begin{figure}[t]
\centering
\includegraphics[width=9cm]{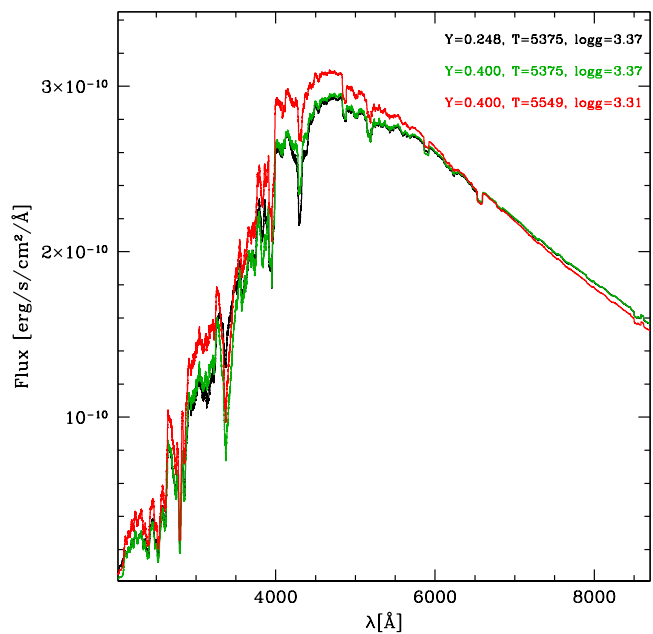}
\caption{Spectral energy distribution of three models with \lL=0.85: one with Y=0.248, \teff\ = 5375 and \logg\ = 3.37 (black line); one with Y=0.400 and the same \teff \ and \logg\ (green line); and the model with Y=0.400 and the associated \teff\ and \logg\ (5549 K and 3.31 respectively, red line).}
\label{comp_sed_L0p85_compo}
\end{figure}

\begin{figure}[t]
\centering
\includegraphics[width=9cm]{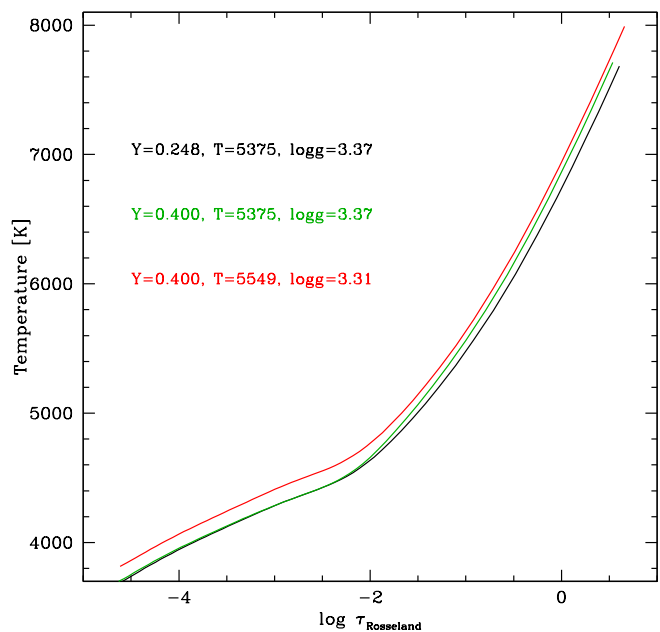}
\caption{Temperature as a function of Rosseland optical depth in the three models leading to the SEDs shown in Fig. \ \ref{comp_sed_L0p85}.}
\label{comp_Tstru_L0p85}
\end{figure}

To further quantify the effect of chemical composition (and in particular of Y) on the SED, in Fig.\ \ref{comp_sed_L0p85_compo} we compare  the SEDs of three models with the same luminosity but different chemical composition. We focus on the models with L=$10^{0.85}$~L$_{\odot}$. We have chosen the models of the Y=0.248 and Y=0.400 sequence, to which we add a third model with the effective temperature and gravity of the Y=0.248 sequence, but the chemical composition of the Y=0.400 sequence. Hence, we are able to disentangle the effect of chemical composition on the effective temperature and logg on one side, and on the SED on the other side. Comparing the two models with \teff\ = 5375 K, we find the same trends as the ones drawn from Fig.\ \ref{comp_CNO}. It is mainly the CH, CN, NH and OH bands that are affected. The general shape of the SED is otherwise relatively similar in the two models. Hence, models with different helium content but similar \teff\ and surface gravity have very similar SEDs, except in regions containing lines involving C, N, and O. If we now compare the two models with the same composition but different effective temperature and surface gravity, we see a major modification of the SED. The hottest model has more (less) flux blueward (redward) of 6000 \AA. This is mainly governed by the change in \teff\ which itself is due to the different helium content. Figure\ \ref{comp_Tstru_L0p85} shows the temperature structure in the three models. Changing only the chemical composition leads to a higher temperature in the inner atmosphere, but a similar structure in the outer parts. This effect is entirely dominated by the difference in the helium content: computing a model with \teff\ = 5375 K, \logg\ = 3.37, and the helium content of the Y=0.400 sequence but the abundances of all elements heavier than He from the Y=0.248 sequence leads to a temperature structure which is almost indistinguishable from that of the green model in Fig.\ \ref{comp_Tstru_L0p85}.  
If we now change both the chemical composition and the effective temperature/surface gravity (red model), we obtain a global increase in the temperature at all depths in the atmosphere, which translates into a change of the SED. 

From these comparisons we conclude that the global shape of the SED is mainly sensitive to the effective temperature which itself depends critically on the helium content and its effect on opacities. The abundances of heavier elements affect certain portions of the SED but not its global shape. Variations in Y thus affect colors sensitive to the global shape of the SED. The filters F395N, F410M, F606W, and F814W should be relatively free of contamination by absorption lines of light elements, and can therefore be used to study the helium content \citep{milone13,milone18,Dotter2015,CassisiSalaris2020}.

%%--------------------------------------------------------------------------
\subsection{Comparison with observed CMDs}
\label{s_compobs}

In this section we compare our synthetic photometry to the observed CMDs of NGC~6752. We use the data of \citet{milone13} and of the HUGS survey \citep{piotto15,nardiello18}. The goal is to see if synthetic colors reproduce observations and to identify potential failures.

\begin{figure*}[]
  \centering
\includegraphics[width=0.4\textwidth]{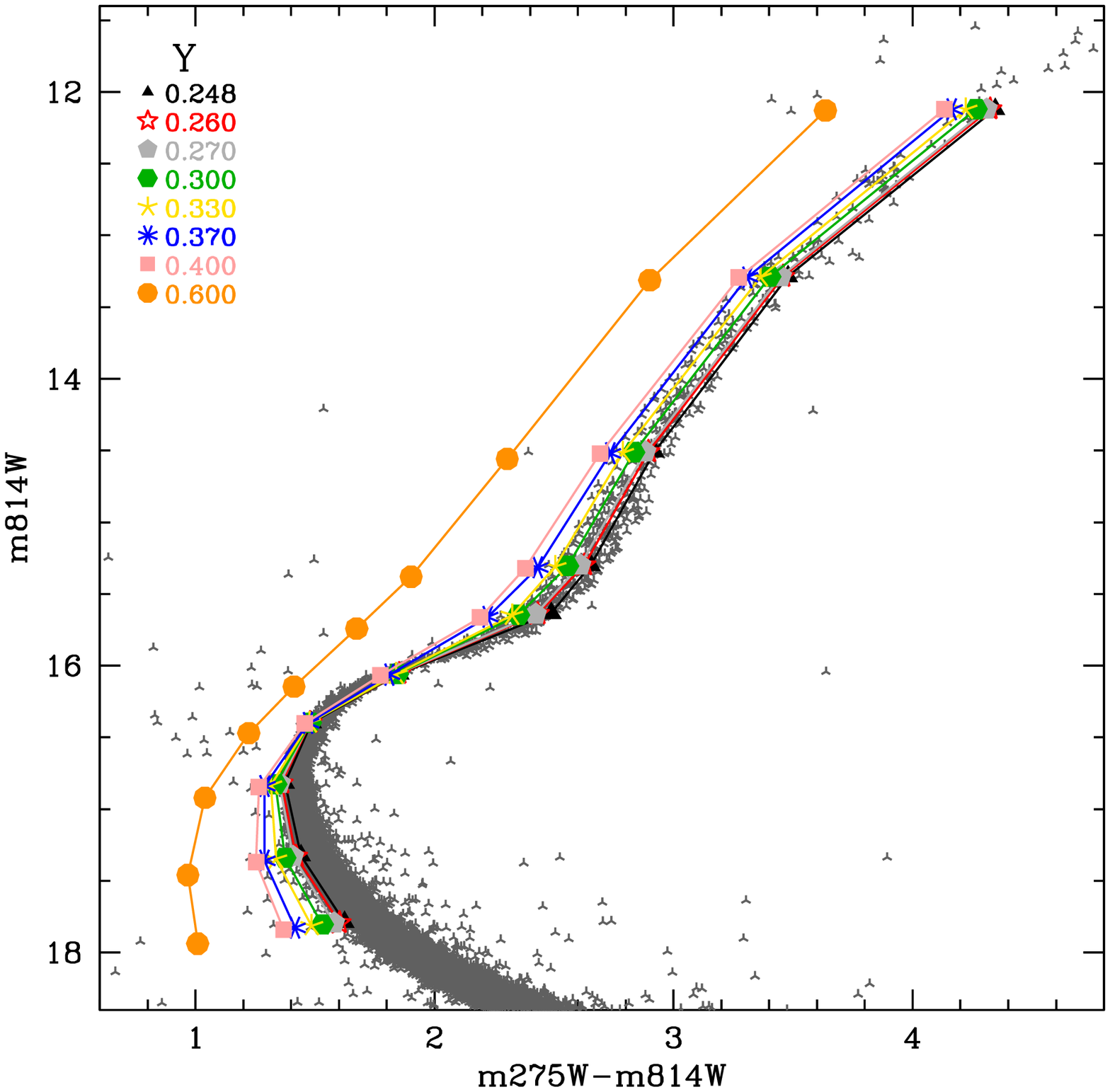}
\includegraphics[width=0.4\textwidth]{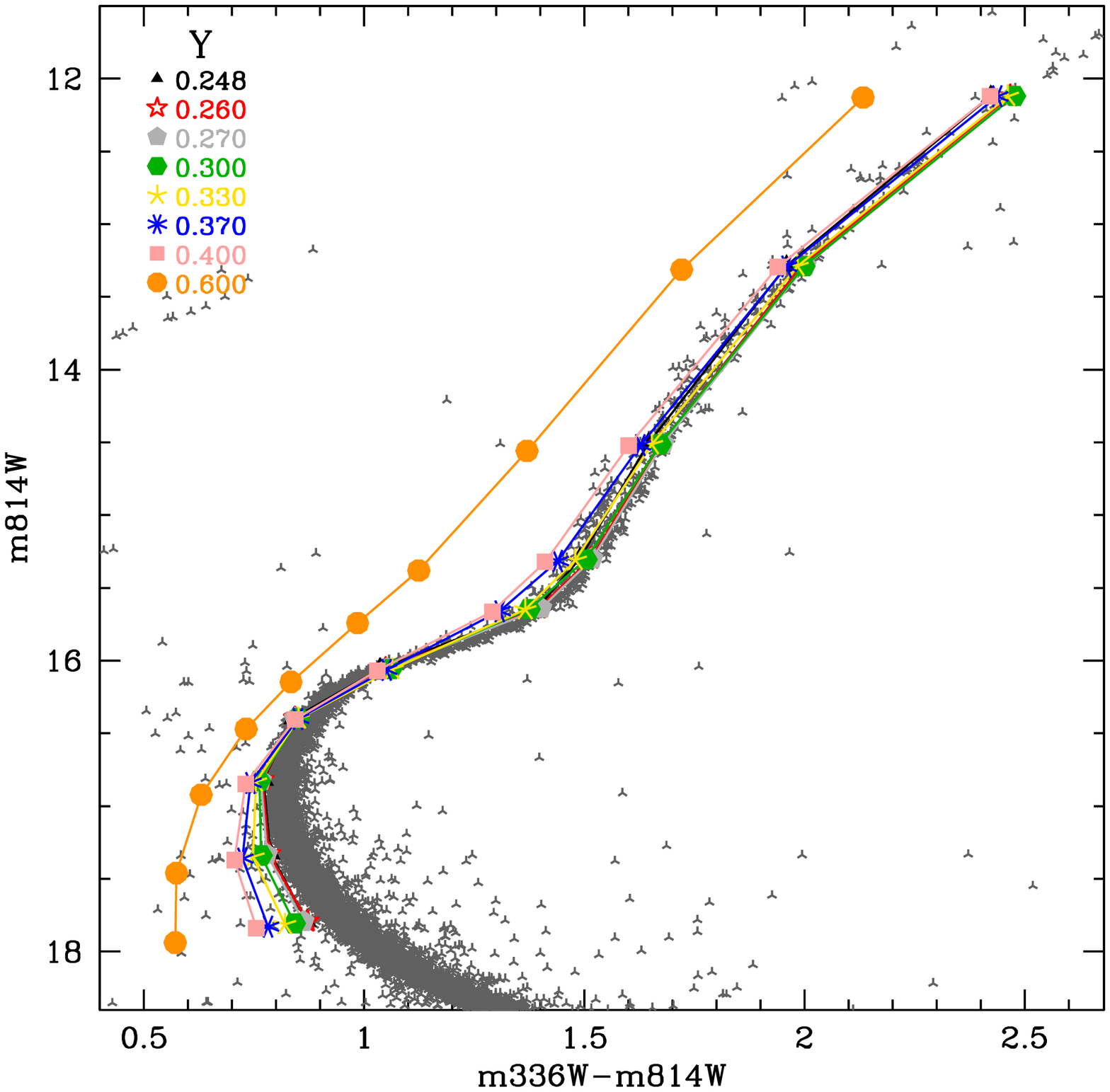}

\includegraphics[width=0.4\textwidth]{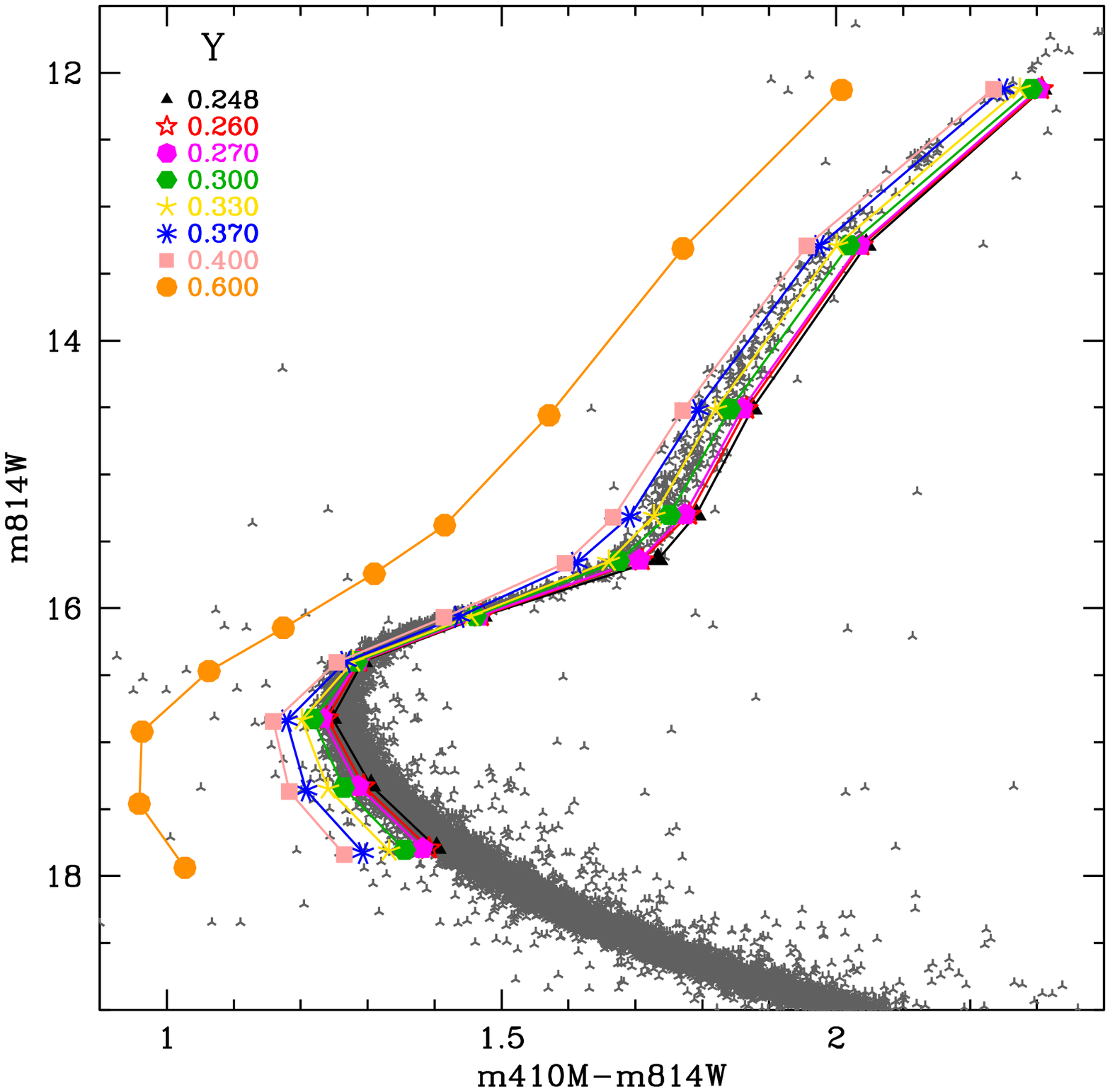}
\includegraphics[width=0.4\textwidth]{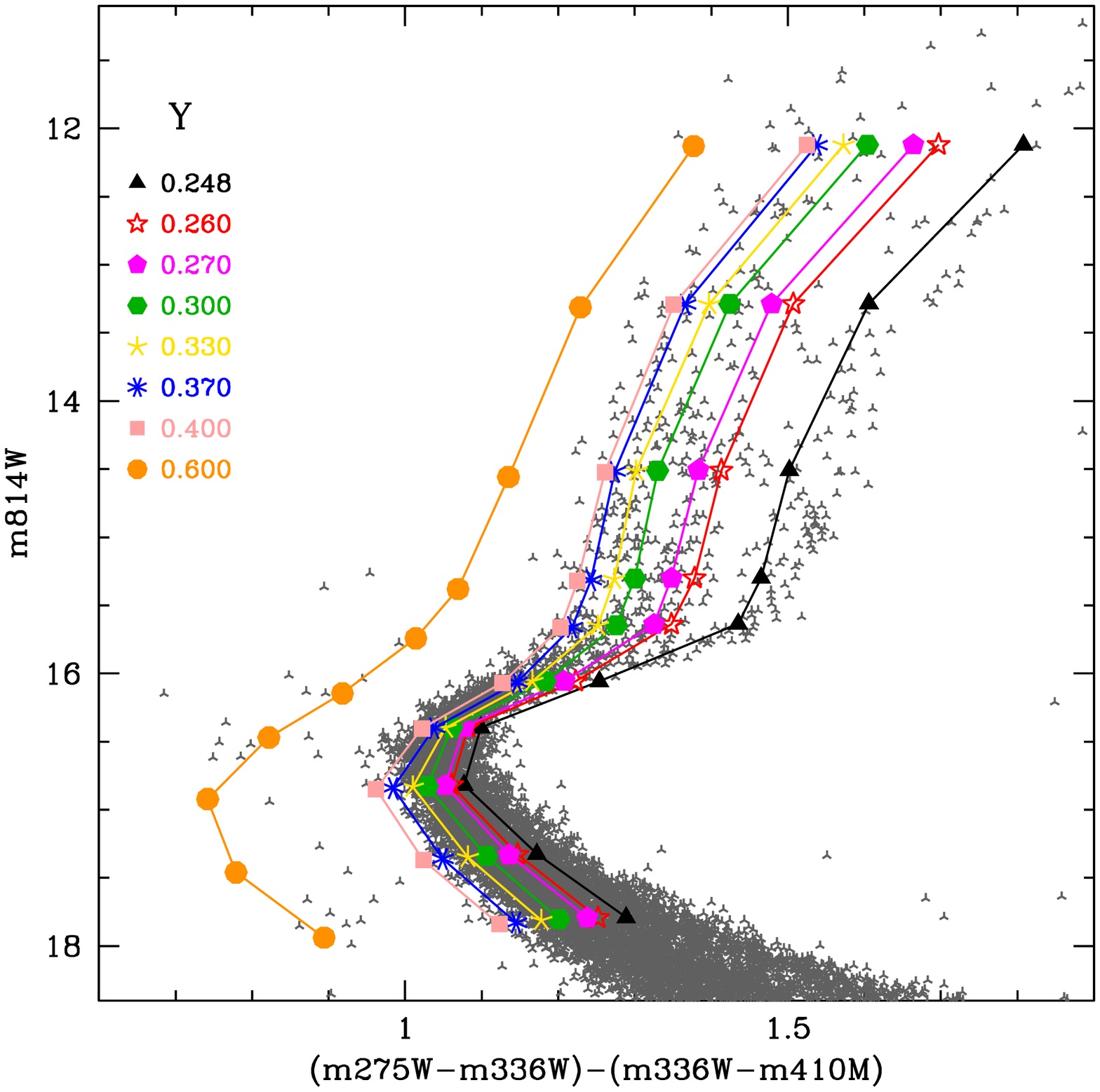}

\includegraphics[width=0.4\textwidth]{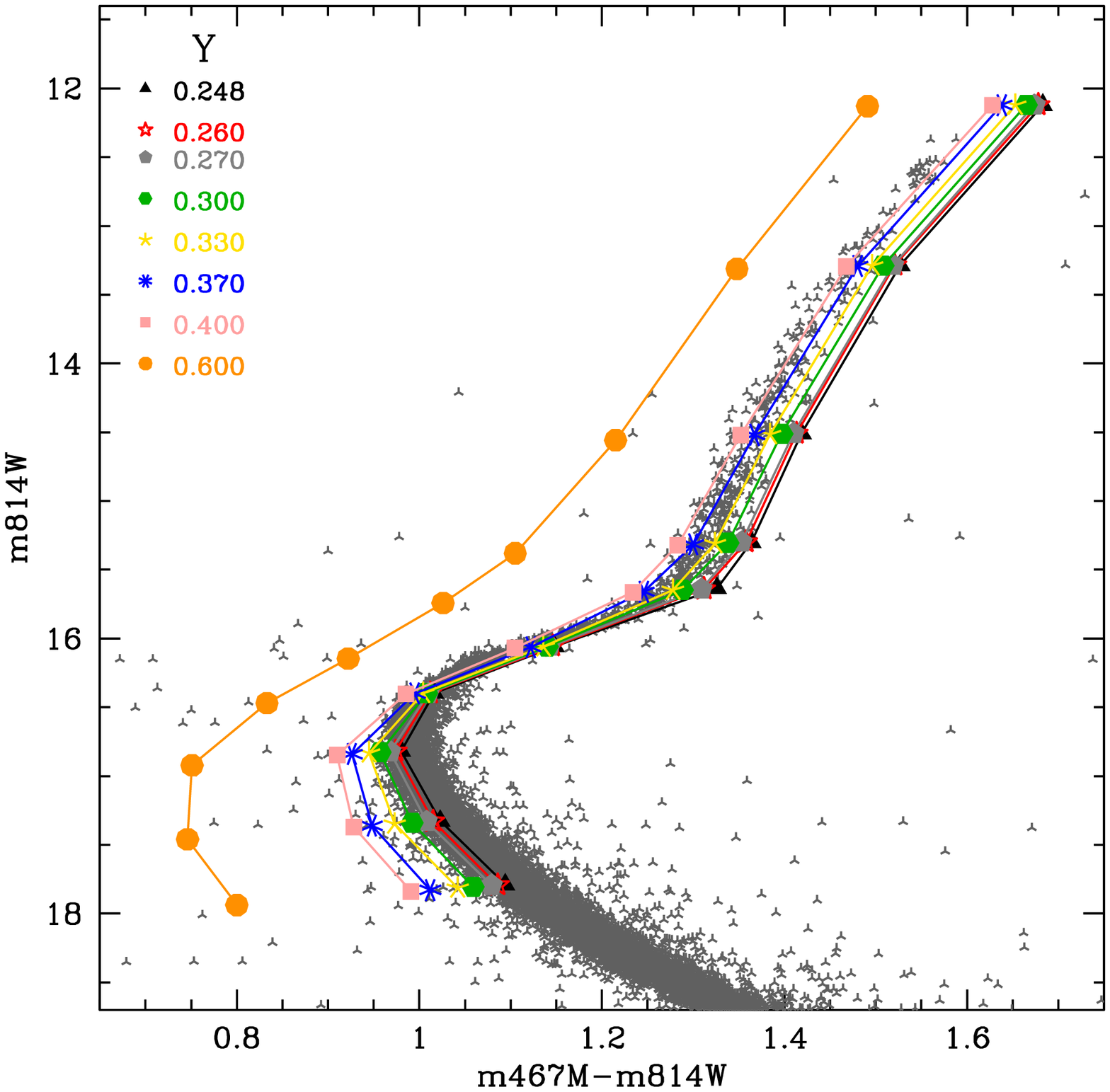}
\includegraphics[width=0.4\textwidth]{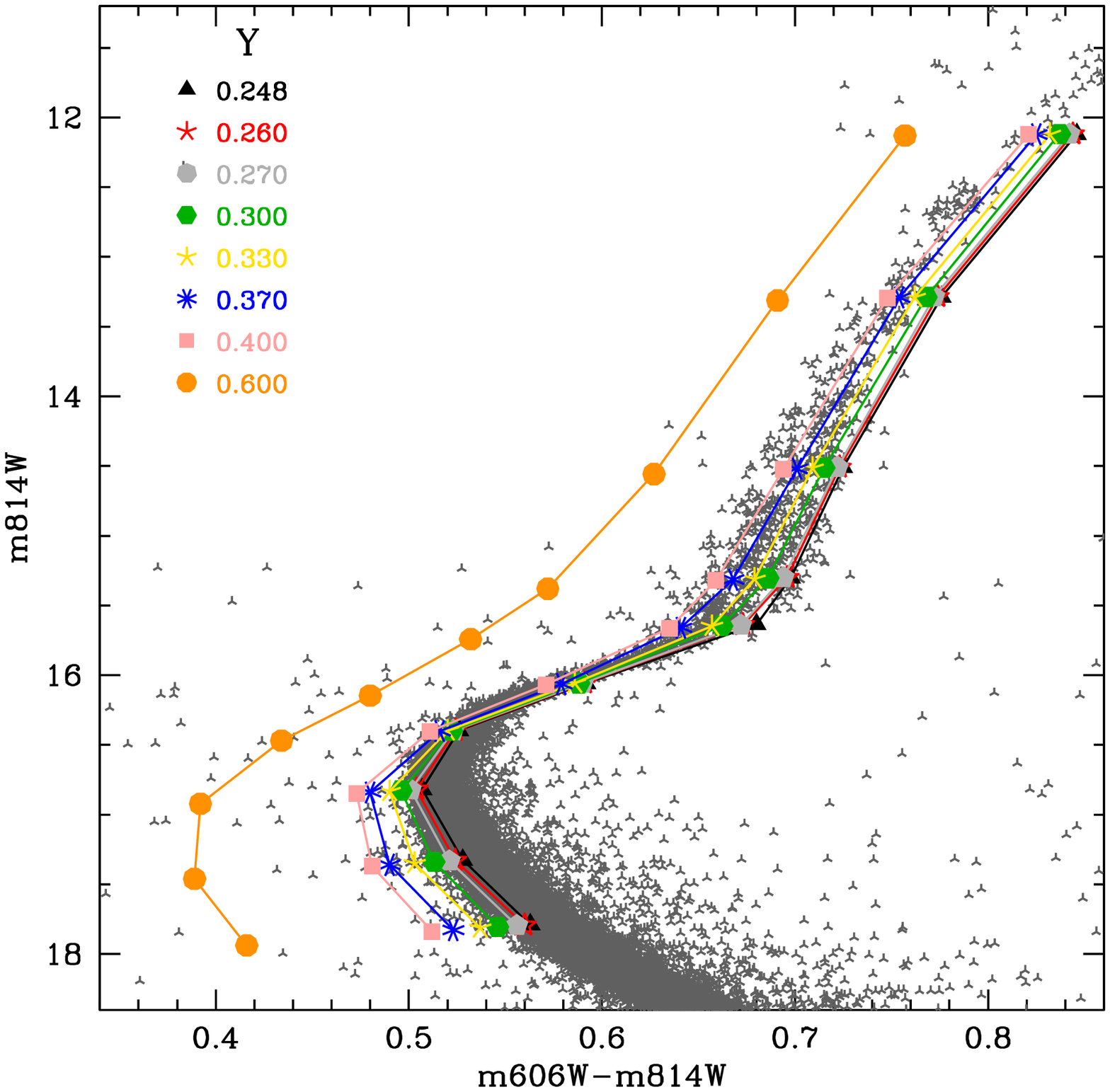} 
\caption{Color-magnitude diagrams: In all panels, the ordinate axis is the magnitude in the ACS F814W filter. The abscissa axis is the difference between magnitude in the X filter and the magnitude in the ACS F814W filter, where X is the WFC3 F275W filter (top left panel), WFC3 F336W filter (top right panel), WFC3 F410M filter (middle left panel), WFC3 F467M filter (bottom left panel), and ACS 606W (bottom right panel). In the middle right panel, the abscissa axis is the color difference (m275W-m336W)-(m336W-m410M). In all panels, different symbols and colors correspond to models with different chemical composition, tagged by their helium content (Y). Gray points correspond to the photometric data of \citet{milone13}, except in the bottom right panel where they are from the HUGS survey \citep{piotto15,nardiello18}.}
\label{cmd}
\end{figure*}

The bottom-right panel of Fig.~\ref{cmd} shows the m814W versus (m606W-m814W) CMD. Taking the Y=0.248 isochrone as a reference, we see that on average our synthetic photometry is able to reproduce the shape of the observed sub-giant branch and the bottom of the RGB. More specifically, the synthetic isochrone is located near the red part of the envelopes that define these branches, which are broadened because of intrinsic dispersion and the presence of multiple populations. The synthetic isochrone reproduces the TO relatively well.
On the upper RGB, the synthetic isochrone appears slightly too red compared to the observations, which could be due to the fact that the stellar evolution models were computed with an atmosphere treated in the gray, plane-parallel, and Eddington approximations.

Figure~\ref{cmd} shows several CMDs involving filters F275W, F336W, F410M, F467M and F814W. These filters are sensitive to the chemical composition as described in Sect.~\ref{s_sed}. 
Focusing on the Y=0.248 isochrone we find, in general, relatively good qualitative agreement with observations. We note that in the m814W versus (m410M-m814W) diagram, the synthetic isochrone is located slightly too much to the red on the RGB (at least when compared to the m814W-(m275W-m814W) and m814W-(m336W-m814W) CMDs). In the (m336W-m814W) color the Y=0.248 isochrone is located roughly in the middle of the observed branch, for reasons that are explained in the following paragraph. The observed RGB is redder than the synthetic isochrone in the m814W versus C$_{410}$ diagram. This is likely the result of the small offsets seen in other CMDs that are amplified by the super color. All these (relatively) small offsets between observations and synthetic isochrones are due to different photometric calibrations between observations and synthetic colors and to limitations in the stellar evolution and spectral modeling. This likely has a small impact on the determination of the helium content of multiple populations, because only relative color differences are used and not absolute ones. However, it is important to keep in mind that full consistency is not achieved between observations and synthetic photometry.

Increasing Y globally leads to a shift of all isochrones to the left because of the increased \teff. An exception is the m814W-(m336W-m814W) CMD where the isochrones move first to the right (Y=0.260 to Y=0.300) and then move back to the blue from Y=0.300  to Y=0.600. This is caused by the strong sensitivity of the F336W filter to the nitrogen abundance (see Sect.~\ref{s_sed}). The N abundance increases rapidly when Y increases, which causes a deepening of the NH absorption band. Consequently, the (m336W-m814W) color is redder. At the same time, \teff\ increases because of the higher helium content. But at first, the effect of nitrogen is stronger, leading to a redder color. When Y reaches $\sim$0.300, the nitrogen increase is not sufficient to counter-balance the effect of the higher \teff\ and the colors become bluer again. This effect, specific to the F336W filter, is not seen in the TO region of the CMD because at the corresponding \teff\ there is no molecular NH band in the  spectra of the stars investigated here.

\vspace{0.5cm}

Having presented and described our synthetic photometry of NGC~6752, we now turn to the main question tackled by this study: the maximum helium content.

%%%%%%%%%%%%%%%%%%%%%%%%%%%%%%%%%%%%%%%%%%%%%%%%%%%%%%%%%%%%%%%%%%%%%%%%%%%%%%%%%%%%%%%%%%%%%%%%%%%%%%%%%%%%%%%%%%%%%%%%%%%%%%%
\section{Maximum helium content}
\label{s_maxHe}

In this section, we investigate whether the current estimates of the maximum helium content of stars of the second population are true values or lower limits. More precisely, we study the possibility that highly enriched stars could be missed when studying multiple populations with HST photometry.

%-------------------------------------------
\subsection{Helium in NGC~6752}
\label{smax_ngc6752}

In a first step, we re-determine the maximum helium mass fraction difference in NGC~6752. We use a  similar method to that of \citet{milone18} which we describe below. Briefly, we select the extreme populations, that is, the least and most chemically enriched ones, from the chromosome map \citep{milone17}. We then estimate the helium mass fraction difference between these two populations in various CMDs using the theoretical isochrones and synthetic photometry presented in Sects.~\ref{s_meth} and \ref{s_sed}.

\begin{figure}[]
\centering
\includegraphics[width=0.49\textwidth]{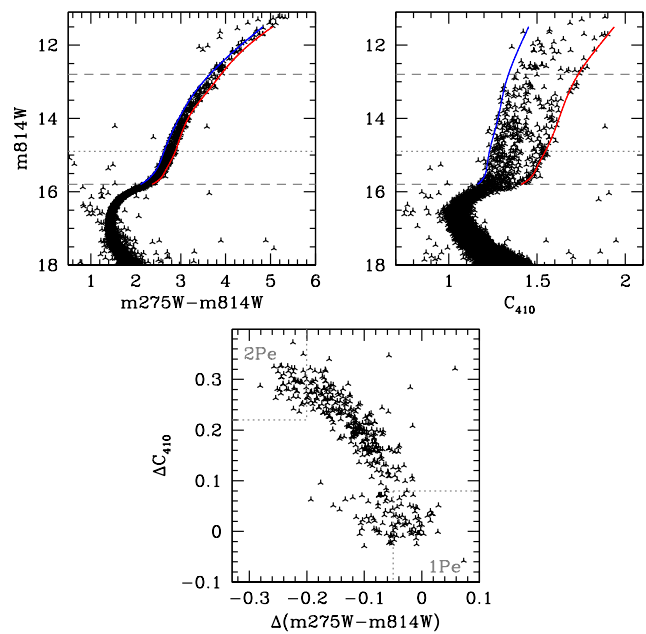}
\caption{\textit{Top left panel}: m814W versus (m275W-m814W) CMD of NGC~6752. \textit{Top right panel}: m814W versus C$_{410}$ diagram. In both panels, the red and blue solid lines are the fiducial lines along the RGB. The horizontal dashed lines indicate the magnitude bin considered to build the chromosome map shown in the bottom panel. The horizontal dotted line marks the m814W value at which the width of the RGB is measured (see text). \textit{Bottom panel}: Chromosome map ($\Delta (C_{410})$ versus $\Delta$ (m275W-m814W)) on the RGB. The dotted lines define boxes in which stars are considered either 1Pe (bottom right) or 2Pe (top left).}
\label{fidchmap6752}
\end{figure}

To build the chromosome map we first create the m814W versus (m275W-m814W) and m814W versus C$_{410}$ diagrams shown in the top panels of Fig.~\ref{fidchmap6752}. We subsequently define the red and blue so-called fiducial lines that bracket the distribution of stars along the RGB. These lines are defined manually by selecting points along the red and blue envelopes and by applying a spline function over the selected points. The width of the RGB is set to the difference between the two fiducial lines at a magnitude of 14.9, corresponding to 2.0 magnitudes above the TO, in agreement with the definition of \citet{milone17}. For each star on the RGB in the  m814W versus (m275W-m814W) and m814W versus C$_{410}$ diagrams, the quantities $\Delta$(m275W-m814W) and $\Delta (C_{410})$ are calculated according to equations 1 and 2 of \citet{milone17}. We select stars with m814W magnitudes between 15.8 and 12.8 (dashed lines in Fig.~\ref{fidchmap6752}) to cover the bottom of the RGB but avoid its part above the bump in the luminosity function where internal mixing can affect surface chemical composition \citep[e.g.,][]{Briley1990,Shetrone2003,CZ2007,Lind2009,Henkel2017}. The resulting chromosome map is shown in the bottom panel of Fig.~\ref{fidchmap6752}.

Following the concept of \citet{milone17}, we manually select the extreme 1Pe and 2Pe stars from the chromosome map by defining boxes respectively near the right-most and left-most parts of the stars' distribution, as illustrated in Fig.~\ref{fidchmap6752}. We then study the color (and thus Y) differences between these two extreme populations in three CMDs:  m814W versus (m395N-m814W),  m814W versus (m410M-m814W), and  m814W versus (m467M-m814W). According to Sect.~\ref{s_compobs} and \citet{milone18} the colors (m395N-m814W), (m410M-m814W), and (m467M-m814W) depend almost exclusively on the helium content at the metallicity of NGC~6752. In addition, the color difference is the largest for the colors that involve filters located at sufficiently blue wavelengths to show sensitivity to effective temperature ---and thus Y--- variations (see Fig.~\ref{comp_sed_L0p85} and middle panel of Fig.~5 of \citealt{milone18}).

Figure~\ref{395m814_1G2G_6752} shows the m814W versus (m395N-m814W) diagram for the 1Pe and 2Pe stars (see Appendix \ref{ap_dY} for the other CMDs). Building on \citet{milone18} we define new fiducial lines along the two populations. We divide each distribution into m814W bins of 0.2 mag  in width. In each bin, we calculate the median m814W and (m395N-m814W). We subsequently perform a boxcar averaging, replacing each median point by the average of its three closest neighbours. This gives the filled circles in Fig.~\ref{395m814_1G2G_6752}. Finally, we run a spline function over these new points to obtain the fiducial lines for each of the two extreme populations. The color difference between these lines is estimated at six m814W magnitudes: 15.2, 14.9, 14.6, 14.3, 14.0, and 13.7. We do not consider brighter stars for reasons that will be explained further below. Section~\ref{ap_dY} of the Appendix shows the CMDs using the F410M and F467M filters.

\begin{figure}[t]
\centering
\includegraphics[width=0.49\textwidth]{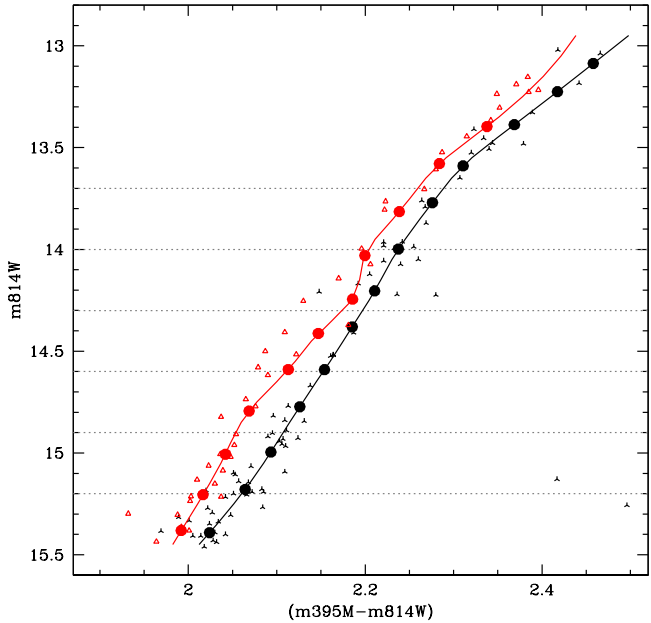}
\caption{m814W versus (m395N-m814W) CMD for the 1Pe (black) and 2Pe (red) populations selected in the chromosome map of Fig.~\ref{fidchmap6752}. The filled circles are the points used to define the fiducial lines, which themselves are shown by the solid lines. See text for details. The horizontal dotted lines indicate the m814W magnitudes at which the color difference between the 2Pe and 1Pe fiducial lines is determined.}
\label{395m814_1G2G_6752}
\end{figure}

\begin{figure}[]
\centering
\includegraphics[width=0.49\textwidth]{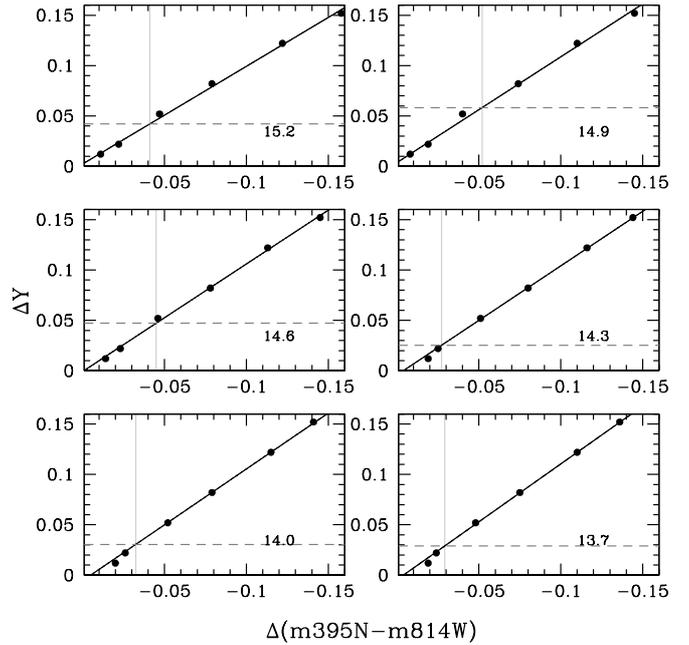}
\caption{Determination of the helium mass fraction difference between 1Pe and 2Pe stars for the six magnitudes (see labels in panels) defined in Fig.~\ref{395m814_1G2G_6752}. For each panel the filled circles are the theoretical values calculated from our isochrones and synthetic photometry. The black solid line is a linear regression to these values. The vertical gray line highlights the measured color difference between 1Pe and 2Pe fiducial lines. The horizontal dashed line indicates the corresponding Y difference read from the black solid line.}
\label{Ydcol395_6752}
\end{figure}

Once obtained, this set of six color differences is compared to theoretical values in Fig.~\ref{Ydcol395_6752}. The latter are calculated from our synthetic photometry, including distance and extinction corrections appropriate for NGC~6752 as described in Sect.~\ref{s_meth}. For each m814W magnitude, we determine the Y difference between the isochrones that match the (m395N-m814W) color difference determined from observations. We do this for the six selected m814W magnitudes and finally average the six determinations to yield the final Y difference. The standard deviation is taken as the uncertainty in this measurement. We perform this process for the three selected colors. The results are gathered in Table~\ref{tab_he_ngc6752}. They are broadly consistent with the determination of \citet{milone18} who quote a maximum Y difference of 0.042$\pm$0.004. We note that our uncertainties are much larger. We tested the effect of varying the size of the boxes to select the 1Pe and 2Pe stars in the chromosome map. Increasing the size of 0.2 magnitudes does not affect the results. However, reducing the size (i.e., selecting fewer points, but at even more extreme positions) translates into an increase in $\Delta$Y by between 0.015 and 0.020. However, at the same time, the uncertainties also increase by the same amount because of
the reduced number of stars,

\begin{table}[ht]
\begin{center}
  \caption{Difference in the helium mass fraction Y between the 1Pe and 2Pe populations, for different colors and for the RGB and MS stars.}
\label{tab_he_ngc6752}
\begin{tabular}{lc}
\hline  
color & $\Delta$Y \\     
\hline
RGB & \\
(m395N-m814W)  & 0.039$\pm$0.013 \\       
(m410M-m814W)  & 0.052$\pm$0.011 \\        
(m467M-m814W)  & 0.068$\pm$0.025 \\
\hline
MS & \\
(m395N-m814W)  & 0.042$\pm$0.004 \\       
(m410M-m814W)  & 0.047$\pm$0.004 \\        
(m467M-m814W)  & 0.049$\pm$0.049 \\
\hline
\end{tabular}
\end{center}
\end{table}

\begin{figure}[]
\centering
\includegraphics[width=0.49\textwidth]{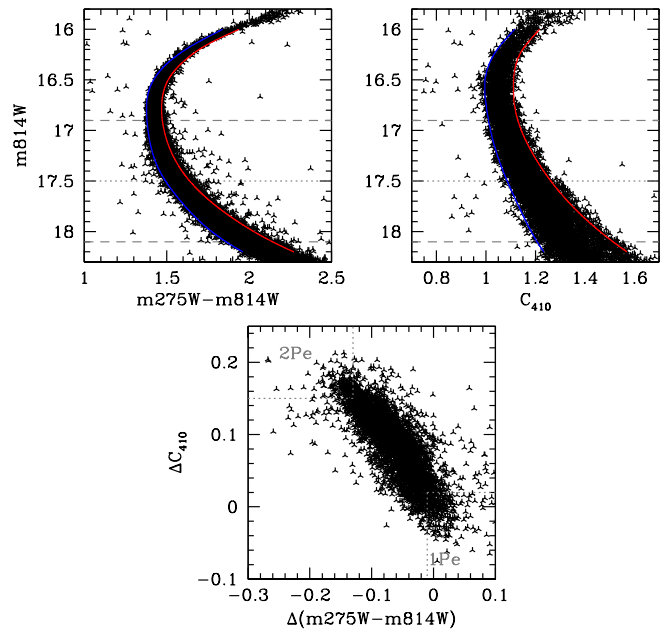}
\caption{Same as Fig.~\ref{fidchmap6752} but for the MS of NGC~6752.}
\label{fidchmap6752MS}
\end{figure}

In anticipation of the following section,  in Fig.~\ref{fidchmap6752MS} and Table~\ref{tab_he_ngc6752} we present the analysis of extreme populations on the MS of NGC~6752. The details of the analysis are presented in Sect.~\ref{clu_W16}. The results indicate that the maximum helium mass fraction reported by \citet{milone18} based on the RGB is also recovered for MS stars. These results are further discussed in Sect.\ref{s_disc}.

%-------------------------------------------
\subsection{Synthetic clusters}
\label{smax_synthclu}

In a second step, we build synthetic clusters to investigate whether or not we underestimate the maximum helium difference between extreme populations. 

\begin{figure}[]
\centering
\includegraphics[width=0.49\textwidth]{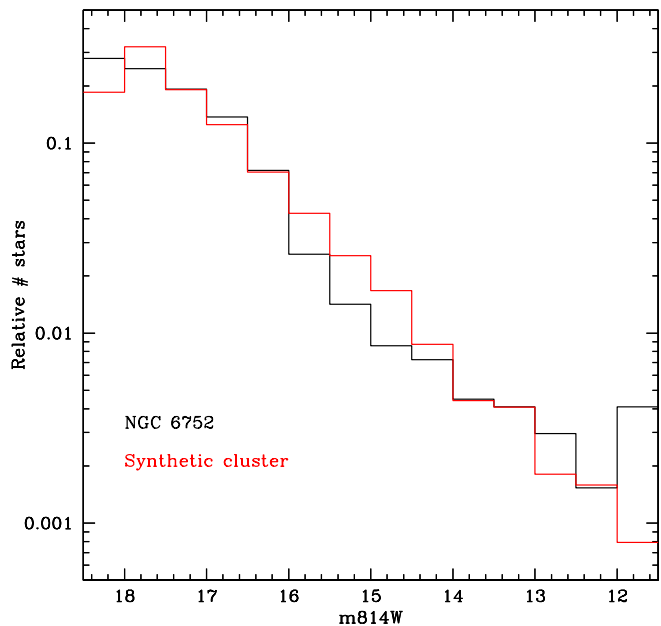}
\caption{Distribution of stars according to their m814W magnitude in NGC~6752 (black) and in a typical synthetic cluster (red). The latter distribution results from random sampling using a Gaussian distribution.}
\label{distrib814}
\end{figure}

We first start by building synthetic CMDs. In practice, we draw artificial points along the isochrones in the m814W versus (m275W-m814W), m814W versus (m336W-m814W), m814W versus (m410M-m814W), m814W versus (m395N-m814W), and m814W versus (m467M-m814W) diagrams (see Fig.~\ref{cmd}). We assume a magnitude distribution in the F814W filter similar to that of NGC~6752 (see Fig.~\ref{distrib814}). We build clusters made of 9000 stars, ensuring a consistent number of stars on the RGB compared to NGC~6752.

For each m814W magnitude drawn along the isochrone (in each diagram) we add a correction to the color (x-axis in each diagram) to introduce a dispersion on the theoretical isochrone. We estimate the correction in the following way. We retrieve photometry of NGC~6752 from the HUGS survey\footnote{\url{https://archive.stsci.edu/prepds/hugs/}} and build the m814W versus (m275W-m814W), m814W versus (m336W-m814W), and m814W versus (m435W-m814W) diagrams. We select stars at m814W magnitudes of 14.9 and 16.5 (in a bin of size 0.2 magnitude). For these subgroups of stars and for each color, (m275W-m814W), (m336W-m814W), and (m435W-m814W), we take the standard deviation with respect to the median point as representative of the dispersion on the RGB and MS. We subsequently use these values to introduce a dispersion in the synthetic colors. For that we randomly draw a color correction by means of a Gaussian distribution centered on zero and characterized by the standard deviation determined above.
In the absence of photometric error for the m395N, m410M, and m467M magnitudes, we assume the dispersion in the colors (m395N-m814W), (m410M-m814W), and (m467M-m814W) is the same as in (m435W-m814W). We perform this process for all isochrones, that is, for all chemical compositions. We finally end up with a series of synthetic CMDs that are used to build synthetic clusters.

%------
\subsubsection{Two chemical compositions}
\label{clu_2pop}

We build synthetic clusters by mixing populations of different chemical composition. We start with a cluster made up of one-third stars with Y=0.248, and two-thirds stars with Y=0.330. For the m814W versus (m275W-m814W) CMD, we therefore select 3000 stars from Y=0.248 isochrone in the synthetic m814W versus (m275W-m814W) just created, and 6000 stars from the Y=0.330 isochrone. We repeat the process for all five diagrams.
We then apply the same method as described in Sect.~\ref{smax_ngc6752} to determine the maximum helium mass fraction difference in the synthetic cluster. We know by construction that it is equal to 0.082. We find that $\Delta$Y=0.081$\pm$0.002, 0.079$\pm$0.002, and 0.070$\pm$0.015 using the (m395N-m814W), (m410M-m814W), and (m467M-m814W) colors, respectively. We thus recover the input value ($\Delta$Y=0.082) with a good level of confidence.

\begin{figure}[]
\centering
\includegraphics[width=0.49\textwidth]{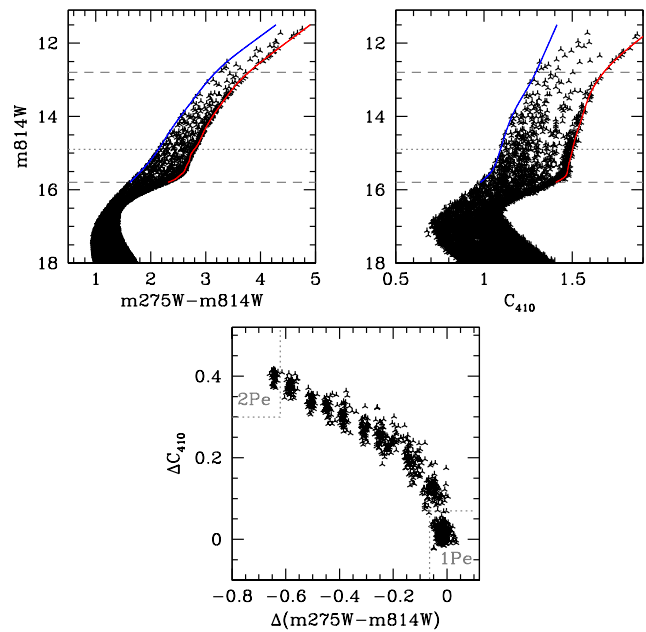}
\caption{Same as Fig.~\ref{fidchmap6752} but for a synthetic cluster with one-third stars with  Y=0.248 and two-thirds stars with Y=0.260 to 0.600 with a flat distribution.}
\label{fidchmapunif0}
\end{figure}

\begin{figure}[]
\centering
\includegraphics[width=0.49\textwidth]{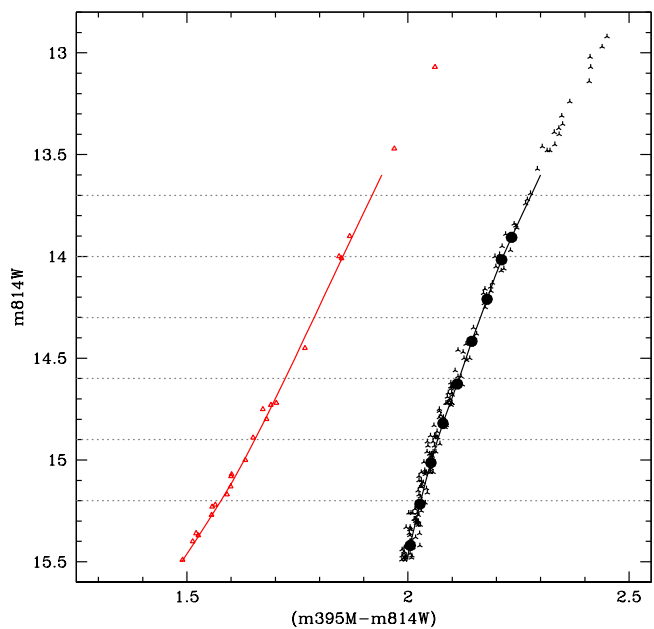}
\caption{Same as Fig.~\ref{395m814_1G2G_6752} but for a synthetic cluster with one-third stars with Y=0.248  and two-thirds stars with Y=0.260 to 0.600 with a flat distribution.}
\label{395m814_1G2G_unif0}
\end{figure}

%------
\subsubsection{Uniform distribution among second population}
\label{clu_unif}

We then run another test by building a synthetic cluster still made up of one-third 1P stars (3000 data points) drawn from the Y=0.248 isochrone, but with the remaining two-thirds of 2P stars (i.e., 6000 data points) spread uniformly between isochrones with Y varying from 0.260 to 0.600. For that, we create the isochrones characterized by Y=0.430, 0.470, 0.500, 0.530, and 0.570 by linearly interpolating between the Y=0.400 and Y=0.600 isochrones. We create the five CMDs with these additional isochrones using the same method as described immediately above. We select 500 stars for each isochrone between 0.260 and 0.600, which is 5.5\% per isochrone, to obtain the final synthetic cluster.
We then proceed as previously to determine $\Delta$Y. Figure~\ref{fidchmapunif0} shows the definition of the fiducial lines and the chromosome map, together with the selected 1Pe and 2Pe stars. Figure~\ref{395m814_1G2G_unif0} shows the 1Pe and 2Pe stars in the m814W versus (m395N-m814W) diagram. Here, we face a problem: the number of 2Pe stars is too small to apply the automatic process for defining the fiducial line of the 2Pe population. We therefore select the fiducial line by hand\footnote{We estimate the uncertainty on $\Delta$Y resulting from the use of this manual process by performing ten repetitions on the same set of 1Pe and 2Pe populations. We find the uncertainty is of the order 0.004.}, as done in the first step of the process that leads to the chromosome map. We then determine the Y difference and we obtain $\Delta$Y=0.362$\pm$0.005. Using the other colors, (m410M-m814W) and (m467M-m814W), we obtain $\Delta$Y=0.364$\pm$0.008 and $\Delta$Y=0.367$\pm$0.007, respectively. The true Y difference, being 0.352, shows that we are able to measure it with good accuracy.

\begin{figure}[]
\centering
\includegraphics[width=0.49\textwidth]{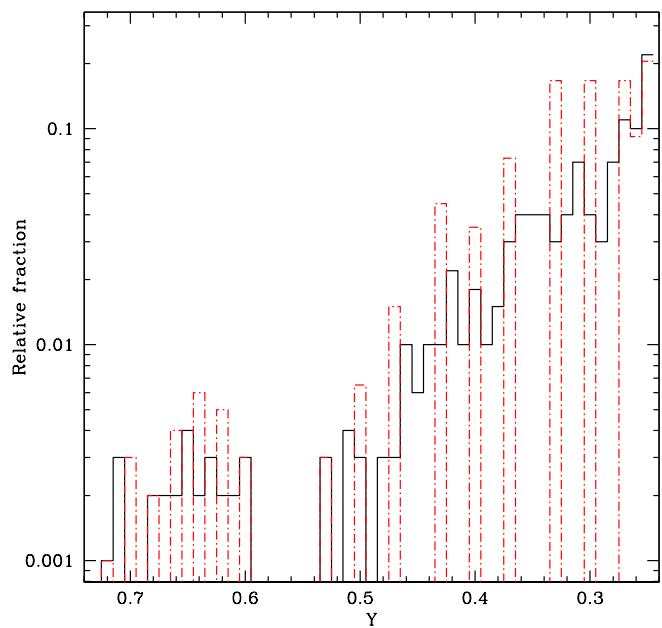}
\caption{Black solid histogram: Relative fraction of stars as a function of their helium content on the lower RGB in the models of \citet{chantereau16} and for an age of 13~Gyr. Red dashed histogram: Distribution we adopt to build synthetic clusters with populations as close as possible to those of \citet{chantereau16}.}
\label{distriW16}
\end{figure}

%------
\subsubsection{Distribution of Chantereau et al.}
\label{clu_W16}

We finally move to clusters made of populations similar to those presented by \citet{chantereau16} within the framework of the FRMS scenario that fits the O-Na anti-correlation in NGC~6752, and which is the most extreme case in terms of He enrichment. We use the distribution of stars according to their helium content shown in Fig.~6 of \citet{chantereau16}  for an age of 13 Gyr (at that age, their model predicts that 10$\%$ of the stars should have Y$>$0.4, from an initial fraction of 21$\%$). We adopt the distribution of the lower RGB as representative of the entire population of the cluster. Examination of Fig.~6 of \citet{chantereau16} reveals that this is a fair approximation for the MS up to the upper RGB. As we do not have synthetic photometry for the full distribution of Y presented by Chantereau et al., we group bins of Y. This is illustrated in Fig.~\ref{distriW16}. For instance,  in the bin with Y=0.30 we gather the stars of Chantereau et al. with Y=0.29, 0.30, and 0.31. In addition, we linearly extrapolated our isochrones beyond Y=0.6 (up to Y=0.72) using our Y=0.4 and Y=0.6 synthetic isochrones. For the populations with Y$>$0.6 in Fig.~\ref{distriW16} we group stars in bins of  0.2 in width (e.g., the Y=0.64 bin gathers stars with Y=0.64 and 0.65 from the distribution of Chantereau et al.). The most enriched population has Y=0.72 in our clusters, corresponding to $\Delta$Y=0.472. We built ten clusters.

\begin{figure}[]
\centering
\includegraphics[width=0.49\textwidth]{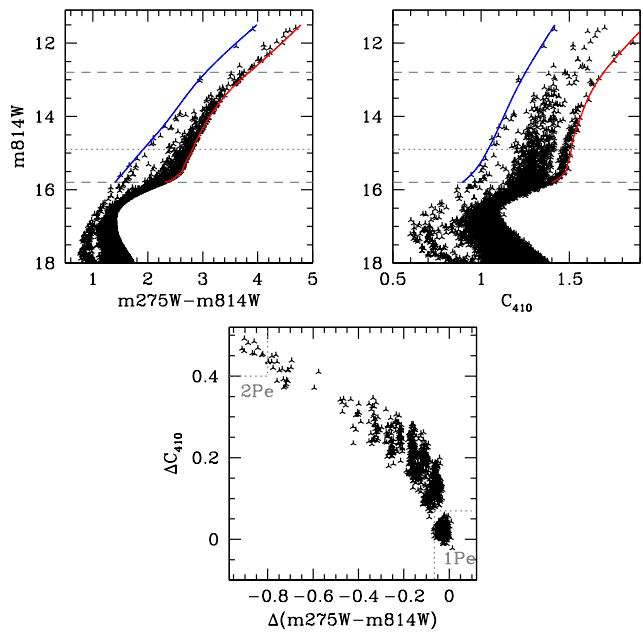}
\caption{Same as Fig.~\ref{fidchmap6752} but for a synthetic cluster with the populations of \citet{chantereau16}.}
\label{fidchmapW16}
\end{figure}

\begin{figure}[]
\centering
\includegraphics[width=0.49\textwidth]{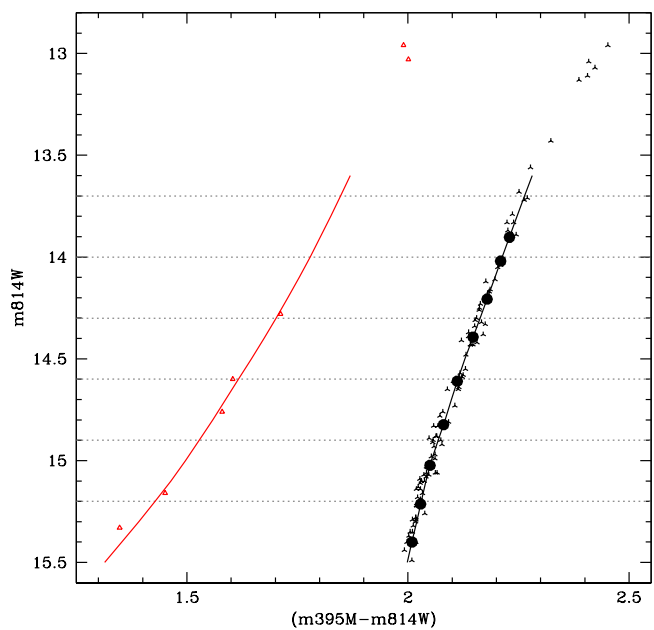}
\caption{Same as Fig.~\ref{395m814_1G2G_6752} but for a synthetic cluster with the populations of \citet{chantereau16}.}
\label{395m814_W16}
\end{figure}

\begin{figure}[]
\centering
\includegraphics[width=0.49\textwidth]{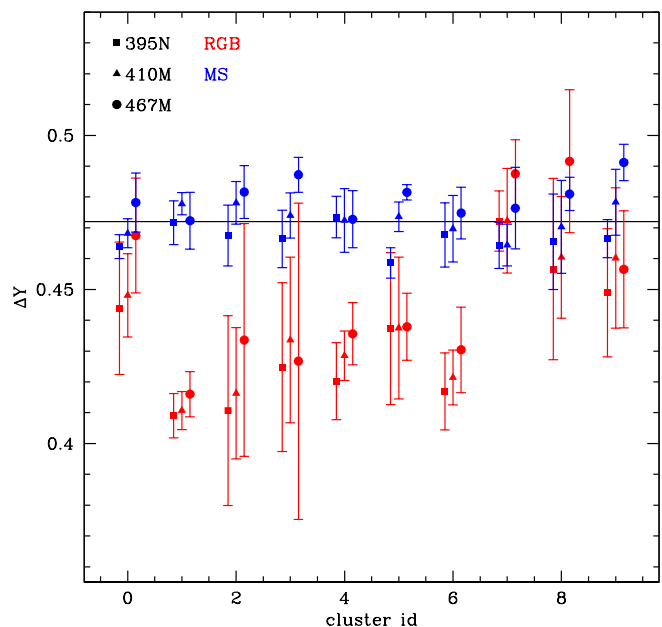}
\caption{Maximum helium difference $\Delta$Y for ten synthetic clusters. Different symbols refer to different colors used for the determination. Blue (red) symbols stand for stars on the MS (RGB). The horizontal black solid line shows the true helium difference used to build the clusters.}
\label{deltaY}
\end{figure}

We then perform the determination of the maximum helium mass fraction as in the previous example where a uniform distribution of stars with Y$>$0.248 was used. Figures~\ref{fidchmapW16} and \ref{395m814_W16} show the synthetic CMDs, chromosome map, and extreme populations in the m814W versus (m395M-m814W) CMD in a representative example. We gather the results in Table~\ref{tab_he_synth} and Fig.~\ref{deltaY}. For some realizations of our synthetic clusters, and for some colors, we derive enrichments that are marginally compatible with the input value $\Delta$Y=0.472: we obtain $\Delta$Y=0.437$\pm$0.038 for cluster 8 using the color (m467M-m814W). However, the main result is that, on average, we determine a maximum $\Delta$Y of about 0.43, which is smaller than the input value of 0.472. This is qualitatively understood by looking at Figs.~\ref{distriW16} and \ref{fidchmapW16}. Stars with Y$>$0.6 have an almost flat distribution which translates into the extended population in upper left part of the chromosome map. When picking the stars identified as the 2Pe population, and to ensure a sufficient number of stars in that population, we include the stars with the highest Y, but also some stars with slightly smaller Y. We therefore tend to create a population with an average Y that is smaller than 0.72.

\begin{table}[t]
\begin{center}
  \caption{Helium difference recovered in ten synthetic clusters with a distribution of the population as in Fig.~\ref{distriW16}. $\Delta$Y$_{X}$ stands for the helium difference estimated from color ($X$-m814W) where $X$ is the magnitude in one of the filters F395N, F410M, or F467M.}
\label{tab_he_synth}
\begin{tabular}{cccc}
\hline  
cluster id & $\Delta$Y$_{395N}$ & $\Delta$Y$_{410M}$ & $\Delta$Y$_{467M}$ \\     
\hline
0  &    0.444$\pm$0.022 &   0.448$\pm$0.014  &   0.468$\pm$0.019 \\
1  &    0.409$\pm$0.007 &   0.411$\pm$0.006  &   0.416$\pm$0.007 \\
2  &    0.411$\pm$0.031 &   0.416$\pm$0.021  &   0.434$\pm$0.038 \\
3  &    0.425$\pm$0.027 &   0.434$\pm$0.027  &   0.427$\pm$0.051 \\
4  &    0.420$\pm$0.013 &   0.429$\pm$0.008  &   0.436$\pm$0.010 \\
5  &    0.437$\pm$0.025 &   0.438$\pm$0.023  &   0.438$\pm$0.011 \\
6  &    0.417$\pm$0.013 &   0.421$\pm$0.009  &   0.430$\pm$0.014 \\
7  &    0.472$\pm$0.010 &   0.472$\pm$0.017  &   0.488$\pm$0.011 \\
8  &    0.457$\pm$0.029 &   0.460$\pm$0.020  &   0.492$\pm$0.023 \\
9  &    0.449$\pm$0.021 &   0.460$\pm$0.023  &   0.457$\pm$0.019 \\
\hline
average & 0.430$\pm$0.004  &  0.426$\pm$0.004  &  0.441$\pm$0.004 \\
\hline
\end{tabular}
\end{center}
\end{table}

\begin{figure}[]
\centering
\includegraphics[width=0.49\textwidth]{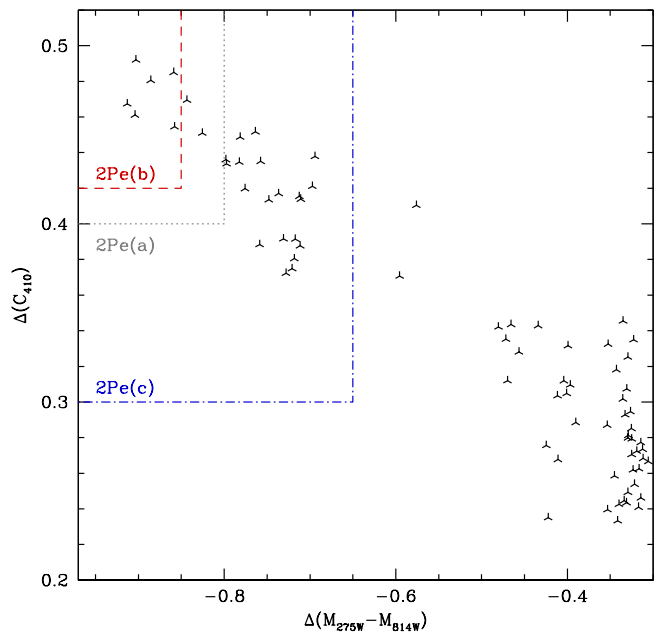}
\caption{Zoom into the upper left part of the chromosome map shown in the bottom panel of Fig.~\ref{fidchmapW16}. The gray dotted lines mark the selection of 2Pe stars as in Fig.~\ref{fidchmapW16}. The blue and red broken lines indicate alternative selections of 2Pe stars.}
\label{chmapzoom2Ge}
\end{figure}

To further investigate this behavior, we show in Fig.~\ref{chmapzoom2Ge} and Table~\ref{tab_he_2Ge} the effect of the selection of 2Pe stars on the determination of the maximum helium difference. Choosing a larger population (population 2Pe(c)) translates into a decrease in $\Delta$Y by $\sim$0.03, as expected because of the inclusion of stars with less extreme Y in the 2Pe population. We also note that the values of $\Delta$Y obtained with that selection of 2Pe stars are consistent with those obtained for other synthetic clusters reported in Table \ref{tab_he_synth} and for which the 2Pe selection was more strict. Inversely, reducing the 2Pe population (2Pe(a) in Fig.~\ref{chmapzoom2Ge}), but still keeping a number of stars sufficient to define the 2Pe fiducial line, leads to a larger $\Delta$Y, which is consistent with the input value.

From these experiments, we conclude that for distributions of multiple populations with small numbers of stars with large chemical enrichments, such as that of \citet{chantereau16}, the derived maximum helium enrichment critically depends on the selection of the 2Pe population. We also find that we slightly underestimate the maximum Y of the cluster.

\begin{table}[]
\begin{center}
  \caption{Effect of the selection of 2Pe stars (see Fig.~\ref{chmapzoom2Ge}) on the determination of the maximum helium difference in synthetic cluster number 0. The first column gives the label of the 2Pe selection. The following columns give the helium difference estimated using the same three colors as in Table~\ref{tab_he_synth}.}
\label{tab_he_2Ge}
\begin{tabular}{cccc}
\hline  
2Pe selection & $\Delta$Y$_{395N}$ & $\Delta$Y$_{410M}$ & $\Delta$Y$_{467M}$ \\     
\hline
2Pe(a)  &       0.444$\pm$0.022 &   0.448$\pm$0.014  &   0.468$\pm$0.019 \\
2Pe(b)  &       0.458$\pm$0.018 &   0.471$\pm$0.011  &   0.489$\pm$0.006 \\
2Pe(c)  &       0.410$\pm$0.040 &   0.410$\pm$0.034  &   0.418$\pm$0.032 \\
\hline
\end{tabular}
\end{center}
\end{table}

\vspace{0.2cm}

\begin{figure}[t]
\centering
\includegraphics[width=0.49\textwidth]{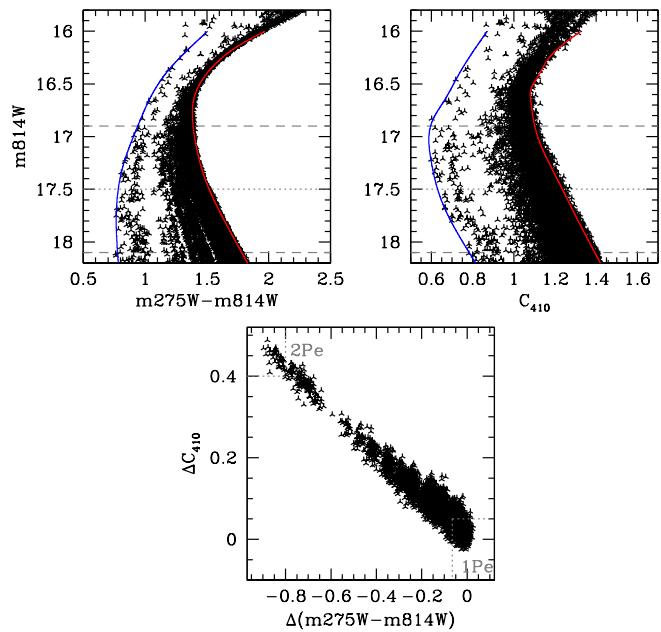}
\caption{Same as Fig.~\ref{fidchmapW16} but for MS stars.}
\label{fidchmapW16_MS}
\end{figure}

\citet{chantereau16}  showed that the RGB contains less very enriched (Y$>$0.6) stars than the MS (see their Fig.~5).
We therefore expect that the determination of the maximum helium content of a GC using MS stars suffers less from the difficulties described above. To test this, we determined $\Delta$Y as reported for RGB stars in Figs.~\ref{fidchmapW16} and \ref{395m814_W16} and Table~\ref{tab_he_synth} but focusing on MS stars. Figure~\ref{fidchmapW16_MS} illustrates our selection of stars: those with m814 magnitudes between 16.9 and 18.1. We estimate the width of the MS at m814=17.5. The 1Pe and 2Pe stars are extracted from the chromosome map as shown in the lower panel of Fig.~\ref{fidchmapW16_MS}. The results are gathered in Table~\ref{tab_he_synth_MS}. Compared to Table~\ref{tab_he_synth} we see that the true helium mass fraction difference (0.472) is indeed better recovered when using MS stars. A graphical representation of this result is given in Fig.~\ref{deltaY} where we see that depending on the cluster simulation, $\Delta$Y based on RGB stars may be underestimated, while for MS stars, the input value is always recovered. We thus conclude that for GCs that would have stellar populations similar to those of \citet{chantereau16}, the study of the maximum helium content should be preferentially performed on MS stars, provided sufficient data quality.

\begin{table}[]
\begin{center}
  \caption{Same as Table~\ref{tab_he_synth} but for MS stars.}
\label{tab_he_synth_MS}
\begin{tabular}{cccc}
\hline  
cluster id & $\Delta$Y$_{395N}$ & $\Delta$Y$_{410M}$ & $\Delta$Y$_{467M}$ \\     
\hline
0  &    0.464$\pm$0.004 &   0.468$\pm$0.005  &    0.478$\pm$0.010 \\
1  &    0.472$\pm$0.007 &   0.478$\pm$0.004  &    0.472$\pm$0.009 \\
2  &    0.468$\pm$0.010 &   0.478$\pm$0.007  &    0.482$\pm$0.009 \\
3  &    0.466$\pm$0.009 &   0.474$\pm$0.007  &    0.487$\pm$0.006 \\
4  &    0.473$\pm$0.007 &   0.472$\pm$0.010  &    0.473$\pm$0.009 \\
5  &    0.459$\pm$0.005 &   0.474$\pm$0.005  &    0.482$\pm$0.003 \\
6  &    0.468$\pm$0.010 &   0.470$\pm$0.011  &    0.475$\pm$0.008 \\
7  &    0.464$\pm$0.008 &   0.464$\pm$0.007  &    0.476$\pm$0.013 \\
8  &    0.466$\pm$0.016 &   0.470$\pm$0.015  &    0.481$\pm$0.005 \\
9  &    0.467$\pm$0.006 &   0.478$\pm$0.011  &    0.491$\pm$0.006 \\
\hline
average & 0.465$\pm$0.002  &  0.473$\pm$0.002  &  0.482$\pm$0.002 \\
\hline
\end{tabular}
\end{center}
\end{table}

%------
\subsubsection{Steep distribution}
\label{clu_steep}

\begin{figure}[]
\centering
\includegraphics[width=0.49\textwidth]{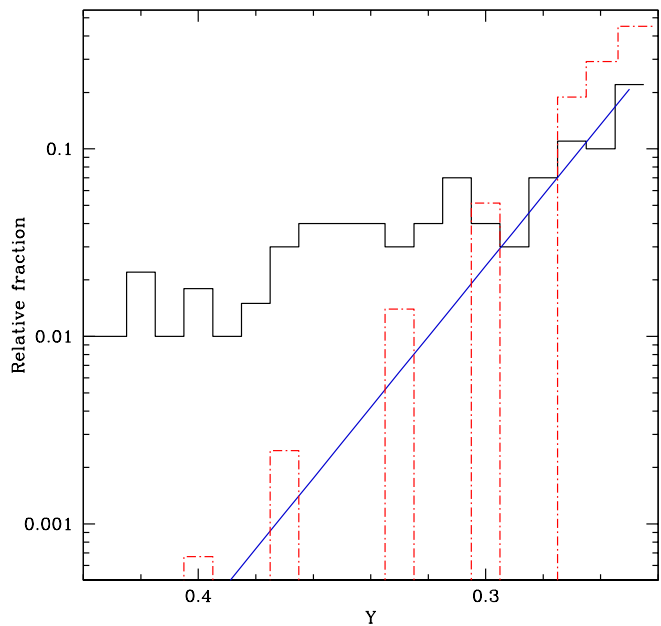}
\caption{Black solid histogram: Relative fraction of stars as a function of their helium content on the lower RGB in the models of \citet{chantereau16} and for an age of 13~Gyr. Blue solid line: Linear interpolation of Chantereau's distribution between Y=0.248 and 0.290, and extrapolation up to Y=0.400. Red dashed histogram: Adopted distribution based on the previous interpolation and re-scaling so that the sum of relative fractions is equal to 1.}
\label{distribsteep}
\end{figure}

In this section we try to answer the question of whether or not we can miss very helium-rich stars if multiple populations follow a steeper distribution than that presented by \citet{chantereau16}. To this end, we define an artificial distribution as shown in Fig.~\ref{distribsteep}. We start from the distribution of \citet{chantereau16} between Y=0.248 and Y=0.29. We linearly interpolate their distribution in this range of Y, and then extrapolate the resulting distribution up to Y=0.400 (blue line in Fig.~\ref{distribsteep}). We thus obtain a much steeper distribution of populations as a function of Y  than that of Chantereau et al. for Y$>$0.29. According to \citet{milone18} and Sect.~\ref{smax_ngc6752}, the maximum helium mass fraction observed in NGC~6752 is about 0.29. Our artificial distribution can therefore be seen as a test case where only a few stars with Y$>$0.29 are present in synthetic clusters. We then try to see if they are recovered or missed when determining $\Delta$Y.

\begin{figure}[]
\centering
\includegraphics[width=0.49\textwidth]{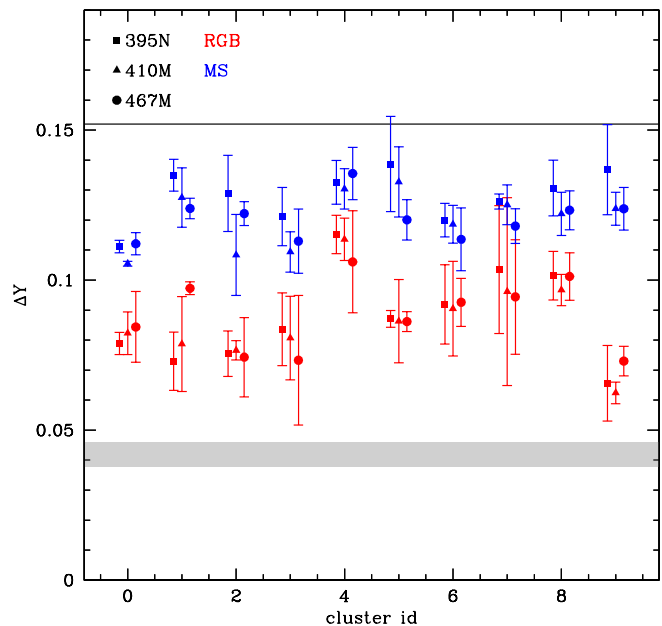}
\caption{Same as Fig.~\ref{deltaY} but for the distribution of the population described in Sect.~\ref{clu_steep}. The gray area marks the observed value of $\Delta$Y in NGC~6752.}
\label{deltaYsteep}
\end{figure}

We built ten synthetic clusters using the above distribution. We use only Y values for which isochrones are available, that is, Y=0.248, 0.260, 0.270, 0.300, 0.330, 0.370, and 0.400 (red bins in Fig.~\ref{distribsteep}). We perform the $\Delta$Y determination as in the previous sections. The results are gathered in Fig.~\ref{deltaYsteep}. Using RGB stars, we are not able to recover the input value ($\Delta$Y=0.152), but we can still detect stars with Y$>$0.29 (i.e., $\Delta$Y=0.042). If present, and under the assumption that their distribution follows that described above, stars more helium-rich than Y=0.29 should therefore be detectable. On the MS, we confirm the trend seen in Sect.~\ref{clu_W16} that higher values of $\Delta$Y are recovered. However, with the present distribution, the maximum helium mass fraction of the cluster is also missed on the MS due to the very small number of highly enriched stars. This is different from Sect.~\ref{clu_W16} where on the MS the initial value of $\Delta$Y could be retrieved.

%%%%%%%%%%%%%%%%%%%%%%%%%%%%%%%%%%%%%%%%%%%%%%%%%%%%%%%%%%%%%%%%%%%%%%%%%%%%%%%%%%%%%%%%%%%%%%%%%%%%%%%%%%%%%%%%%%%%%%%%%%%%%%%
\section{Discussion}
\label{s_disc}

Our method for determining the maximum helium mass fraction in GCs is slightly less automated than that presented by \citet{milone18}. In particular, the selection of the extreme 1Pe and 2Pe populations is made visually in our case, while it relies on a more sophisticated treatment in \citet{milone18}. We tested the effect of the 2Pe selection in Sect.~\ref{clu_W16} and Fig.~\ref{chmapzoom2Ge}, and argue that the choice of 2Pe stars can affect the value of the maximum Y. However, qualitatively, stars much more He-rich than the observed limit of 0.042 are easily detected regardless of the 2Pe selection. In addition, the automated method of \citet{milone18} works well when 2Pe stars are clearly identified by an overdensity in the chromosome map, but there are clusters for which no such overdensity is visible. For instance, the chromosome map of NGC~6752 itself, presented in the bottom panel of Fig.~\ref{fidchmap6752}, shows a rather uniform distribution along the 2G sequence (i.e., from the bottom right corner up to the upper left corner). Hence, we argue that the choice of 2Pe stars in our study is not drastically different from \citet{milone18}.

He-rich populations may be contaminated by blue stragglers and their descendants which would populate similar regions of CMDs. \citet{marino19} showed that candidate evolved blue stragglers should be located preferentially along the 1P sequence, and thus would not contaminate 2Pe stars. However, one may wonder whether blue stragglers resulting from the merger of two 2P stars could produce stars along the 2P sequence too. This needs to be investigated further.  
Binaries can also contaminate the 2P sequence. As shown by \citet{martins20}, the presence of a companion among RGB stars will tend to displace a star up and left in the chromosome map. The magnitude of the displacement depends on the relative brightness and effective temperature of the two stars. Hence, stars along the 2G sequence, with medium Y but with a companion, may be moved to the position of single stars with higher Y. Consequently, the extremity of the 2G sequence may contain stars with less extreme Y. This could lead to overestimation of the maximum Y. The binary fraction among GCs is low, usually below 10\% \citep{sollima07,milone12,jb15}, and their effect on the maximum Y should therefore be limited; unless the number of very He-rich stars is also small, as in the distribution of \citet{chantereau16}. In any case, if binaries are present, the maximum Y determined from photometry is probably overestimated.

With these limitations in mind, and given the results of the present study, it is unlikely that  NGC~6752 contains stars with Y$\gtrsim$0.3. If multiple populations were formed as predicted by the FRMS scenario developed by \citet{decressin07} and \citet{chantereau15,chantereau16}, a wide distribution of Y, from 0.248 to 0.72, should be present among stars either on the MS or the RGB, with ~39$\%$ of the stars with Y$>$0.3 and ~10$\%$ with Y$>$0.4 at 13~Gyr. Our study reveals that while we may not retrieve the most helium-rich objects on the RGB, we should still be able to detect stars with Y as high as 0.65 (see Fig.~\ref{deltaY}). This is clearly in contrast with results based on HST photometry indicating Y no higher than 0.3. We have shown that even in the case of a Y distribution much steeper than that predicted by \citet{chantereau16} we would be able to find stars with helium enrichment beyond the observed value.

A prediction of our study dedicated to NGC~6752 is that the maximum helium content of GC stars, if it follows a distribution where helium-rich stars are less numerous than helium-poor ones, is best determined on the MS rather than on the RGB. This is due to the faster evolution of He-rich stars and consequently the larger number of stars on the MS than on the RGB \citep{dm73,dantona10,chantereau16}. 
However, the maximum Y we obtain on the MS using HST data of NGC~6752 does not indicate a significant difference compared to the value found on the RGB.

All in all, the present results strongly suggest that stars in NGC~6752 do not follow the distribution predicted by the FRMS model presented in \citet{chantereau16}, and consequently, multiple populations were not likely formed out of material polluted by this type of object.
This conclusion applies only to NGC~6752, and additional studies of clusters with different ages, metallicities, and masses are required to see whether generalizations can be made. A wider study of this kind is necessary to investigate whether or not the current observational limit of Y$\sim$0.4 is robust and to investigate whether or not it suffers from observational limitations. As recalled in Sect.~\ref{s_intro}, this limit is a key prediction of some models, but also a building block of some others. In particular, the formation of multiple populations caused by pollution of material ejected from a super-massive star (SMS) originally {assumes} that the SMS stops its evolution when its core helium content reaches 0.4, and is further dislocated by instabilities and/or strong stellar winds \citep{denis14}. This assumption was dictated  by the observational fact that Y appears to be no higher than 0.4 in the most extreme GC populations. On the other hand, hot-hydrogen burning products with low helium abundances in agreement with the photometric constraints can be ejected by SMSs in the case where they are continuously rejuvenated by runaway stellar collisions \citep{Gieles2018}.

Pollution by material produced in massive AGB stars predicts that the helium distribution among multiple populations reaches a maximum of $\sim$0.38 \citep[e.g.,][]{dercole10}. This is consistent with the current observational limit, and would make these objects good polluter candidates if other problems were not linked to their use, as summarized in \citet{Renzini2015} and \citet{bastianlardo18} for instance. Other important factors are the time constraints on the dilution by pristine material, as also highlighted by \citet[][see also \citealt{dercole16}]{dercole11}, and the mass budget issue \citep[e.g.,][]{PC2006,Krause2016}. The transformation of the Na-O correlation in AGB ejecta into a Na-O anti-correlation, as seen in all GCs, also requires specific conditions for mixing of pristine material with nucleosynthesis products \citep[e.g.,][]{Ventura2008,Karakas2010}. In addition, the C+N+O sum of AGB ejecta is not conserved \citep{Decressin2009}, contrary to what is deduced from spectroscopic analysis of stars in NGC~6752 \citep{yong15}. 

The role of massive binaries in the multiple population phenomenon remains to be investigated. \citet{demink09}  showed that these objects could potentially produce the required abundance patterns. But their study is limited to a 20+15 \msun\ system on a 12-day orbit and in which both components have reached synchronization. The average helium mass fraction of the ejecta of this binary system is 0.3. Material ejected at later phases of the system's evolution has Y as large as $\sim$0.63 (see Fig.~1 of de Mink et al.). The average helium content is therefore consistent with the current maximum Y observed in GCs. However binary evolution depends not only on the properties and evolution of the  components, but also on the parameters of the  system (separation, eccentricity, mass ratio, mass transfer efficiency); see for example \citet{menon20}. Additionally, the maximum central temperature of the considered mass domain does not reach the high values required to build the Mg-Al anti-correlation \citep[e.g.,][]{prantzos07,Prantzos2017}. A wider study involving population synthesis is therefore required to assess the impact of massive binaries on the origin of multiple populations in GCs.

%%%%%%%%%%%%%%%%%%%%%%%%%%%%%%%%%%%%%%%%%%%%%%%%%%%%%%%%%%%%%%%%%%%%%%%%%%%%%%%%%%%%%%%%%%%%%%%%%%%%%%%%%%%%%%%%%%%%%%%%%%%%%%%
\section{Conclusion}
\label{s_conc}

In this study, we investigated the determination of the maximum helium mass fraction in stars of the GC NGC~6752. Our goal was to decipher whether we really detect the most He-rich stars with present-day
photometric methods, or miss them.
We relied on the work of \citet{chantereau16} who produced isochrones with various chemical compositions corresponding to different degrees of pollution by FRMS which is an extreme case in terms of He enrichment. We computed synthetic spectra along these isochrones using the atmosphere code ATLAS12 and the spectral synthesis code SYNTHE. The resulting spectra were used to compute synthetic photometry in the following HST filters: WFC3 F275W, WFC3 F336W, WFC3 F410M, WFC3 F467M, ACS F606W, and ACS F814W. We compared the synthetic colors with data of NGC~6752 obtained by \citet{milone13}. The different CMDs are usually reasonably well reproduced, although offsets exist between synthetic and observed sequences (MS, TO, RGB).

We re-determined the maximum helium mass fraction of stars in NGC~6752 using a method very similar to that of \citet{milone18}. Our results are consistent with those of Milone at al.
We built synthetic clusters with various populations characterized by their He content (they also have different composition in light elements). We validated our method on simple population distributions, ensuring that we are able to recover the input maximum Y. We subsequently created synthetic clusters following the distribution presented by \citet{chantereau16}, that is, with stars with Y of between 0.248 and 0.72. We show that on the RGB, we slightly underestimate the maximum Y, but by a relatively small amount ($\sim$0.05). On the MS, we retrieve the input value. In any case, even on the RGB the maximum Y we determine in synthetic clusters is higher than the observed value of 0.042$\pm$0.004 \citep{milone18}. We tested that even if populations followed a steeper Y distribution than that of \citet{chantereau16}, stars with Y higher than 0.042 are recovered (in that case the maximum Y is underestimated both on the RGB and MS). Finally, we determined the maximum Y on the MS of NGC~6752 using HST data and find it to be consistent with the value obtained on the RGB.

These results indicate that multiple populations in NGC~6752 have a Y distribution that is very likely not that assumed by \citet{chantereau16} in the framework of the FRMS scenario. Our results also show that in the specific case studied here, the maximum helium mass fraction determined from observations is probably the true value (i.e., there are no stars more He-rich in that cluster). 
We stress that although we have focused on the specific case of pollution by FRMS, our results apply to any scenario that would predict a strong He enrichment among 2P stars \citep{Salarisetal2006,Pietrinferni2009,Cassisi2013,Dotter2015}.
Our results need to be extended to other clusters spanning a range of parameters (age, metallicity, mass) in order to confirm that FRMSs are not responsible for the multiple populations in GCs and to better constrain the scenario that produce them.

%%#####################################################################
\section*{Acknowledgments}

We thank an anonymous referee for a positive report. We thank A. Milone for sharing HST photometry of NGC~6752 and for interesting discussions. This work was supported by the Swiss National Science Foundation (Project 200020-192039 PI C.C.). This work was supported by the Agence Nationale de la Recherche grant POPSYCLE number ANR-19-CE31-0022.

%%#####################################################################
\bibliographystyle{aa}
\bibliography{clu_issi}
%%#####################################################################
%%#####################################################################

%\newpage

%%%%%%%%%%%%%%%%%%%%%%%%%%%%%%%%%%%%%%%%%%%%%%%%%%%%%%%%%%%%%%%%%%%%%%%%%%%%%%%%%%%%%%%%%%%%%%%%%%%%%%%%%%%%%%%%%%%%%%%%%%%%%%%
\begin{appendix}

%*******************
\section{Chemical composition adopted for multiple populations}
\label{ap_chem}

Table \ref{tab_chem} gives the chemical composition of the different mixtures used to compute evolutionary tracks and isochrones according to \citet{chantereau15} adapted for [Fe/H] = -1.53. Each mixture is labeled according to its helium mass fraction Y. The Y=0.248 mixture is that of unpolluted stars, while all the others corresponds to various degrees of pollution by FRMS \citep[see][]{chantereau15}. We stress that on the giant branches the chemical compositions are additionally affected by stellar evolution. We therefore give the initial composition that is encountered on the MS.

\begin{table*}[ht]
\begin{center}
  \caption{Mass fraction of key elements on the MS for several chemical compositions labeled by their helium mass fraction Y.}
\label{tab_chem}
\begin{tabular}{lcccccccc}
\hline
Specie & Y=0.248       &  Y=0.260      & Y=0.270      & Y=0.300       & Y=0.330      & Y=0.370        & Y=0.400     & Y=0.600 \\    
\hline
H      & 0.751         & 0.739         & 0.729        & 0.699         & 0.669         & 0.629        & 0.599         & 0.399 \\
He     & 0.248         & 0.260         & 0.270        & 0.300         & 0.330         & 0.370        & 0.400         & 0.600 \\
C      & 9.4 10$^{-5}$  & 8.6 10$^{-5}$  & 8.0 10$^{-5}$ & 7.0 10$^{-5}$  & 6.2 10$^{-5}$  & 5.5 10$^{-5}$ & 5.1 10$^{-5}$ & 3.9 10$^{-5}$ \\
N      & 2.9 10$^{-5}$  & 8.9 10$^{-5}$  & 1.2 10$^{-4}$ & 1.8 10$^{-4}$  & 2.3 10$^{-4}$  & 2.8 10$^{-4}$ & 3.0 10$^{-4}$ & 4.0 10$^{-4}$ \\
O      & 5.2 10$^{-4}$  & 4.7 10$^{-4}$  & 4.4 10$^{-4}$ & 3.8 10$^{-4}$  & 3.4 10$^{-4}$  & 2.9 10$^{-4}$ & 2.7 10$^{-4}$ & 1.7 10$^{-4}$\\
Na     & 1.1 10$^{-6}$  & 2.7 10$^{-6}$  & 3.3 10$^{-6}$ & 4.8 10$^{-6}$  & 5.7 10$^{-6}$  & 6.6 10$^{-6}$ & 7.1 10$^{-6}$ & 7.7 10$^{-6}$\\
Mg     & 3.6 10$^{-5}$  & 3.6 10$^{-5}$  & 3.6 10$^{-5}$ & 3.5 10$^{-5}$  & 3.5 10$^{-5}$  & 3.5 10$^{-5}$ & 3.4 10$^{-5}$ & 3.3 10$^{-5}$\\
Al     & 1.8 10$^{-6}$  & 2.4 10$^{-6}$  & 2.8 10$^{-6}$ & 3.7 10$^{-6}$  & 4.4 10$^{-6}$  & 5.2 10$^{-6}$ & 5.6 10$^{-6}$ & 7.4 10$^{-6}$\\
\hline
\end{tabular}
\end{center}
\end{table*}

%*******************
\section{Y difference from various colors}
\label{ap_dY}

In this Appendix we show the determination of the maximum Y difference in NGC~6752 using the colors (m410M-m814W) and (m467M-m814W).

\begin{figure*}[]
\centering
\includegraphics[width=0.49\textwidth]{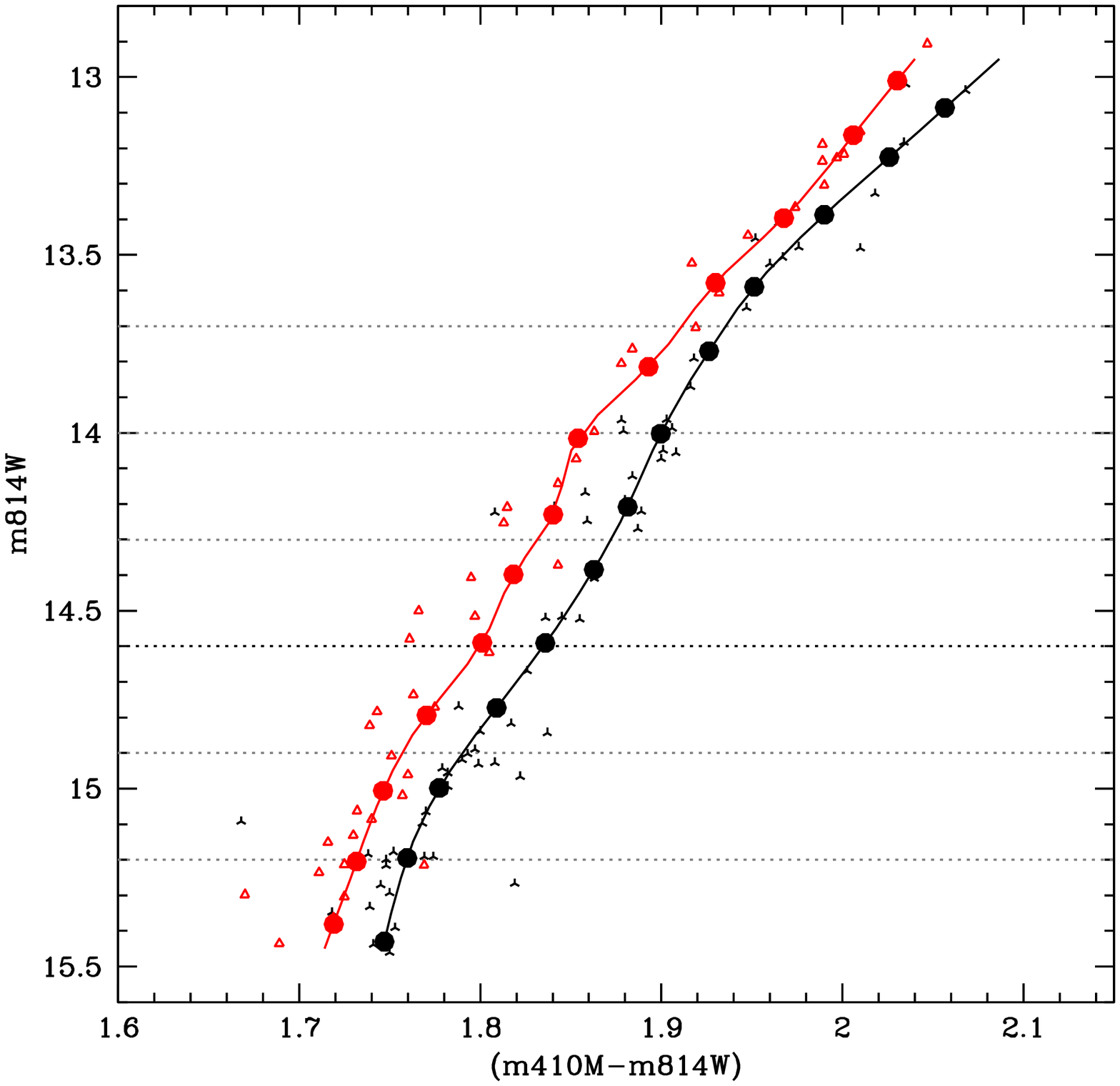}
\includegraphics[width=0.49\textwidth]{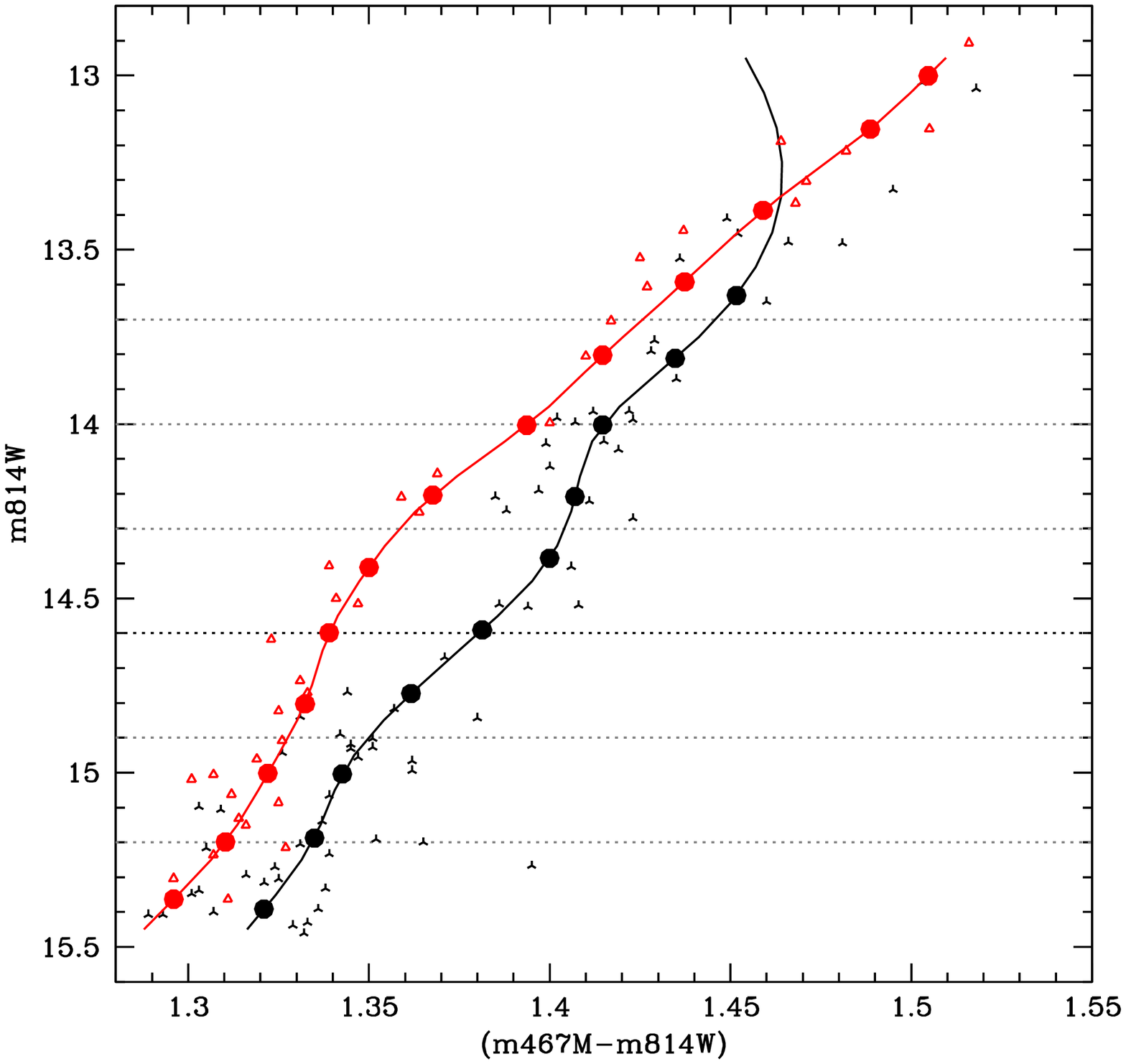}

\includegraphics[width=0.49\textwidth]{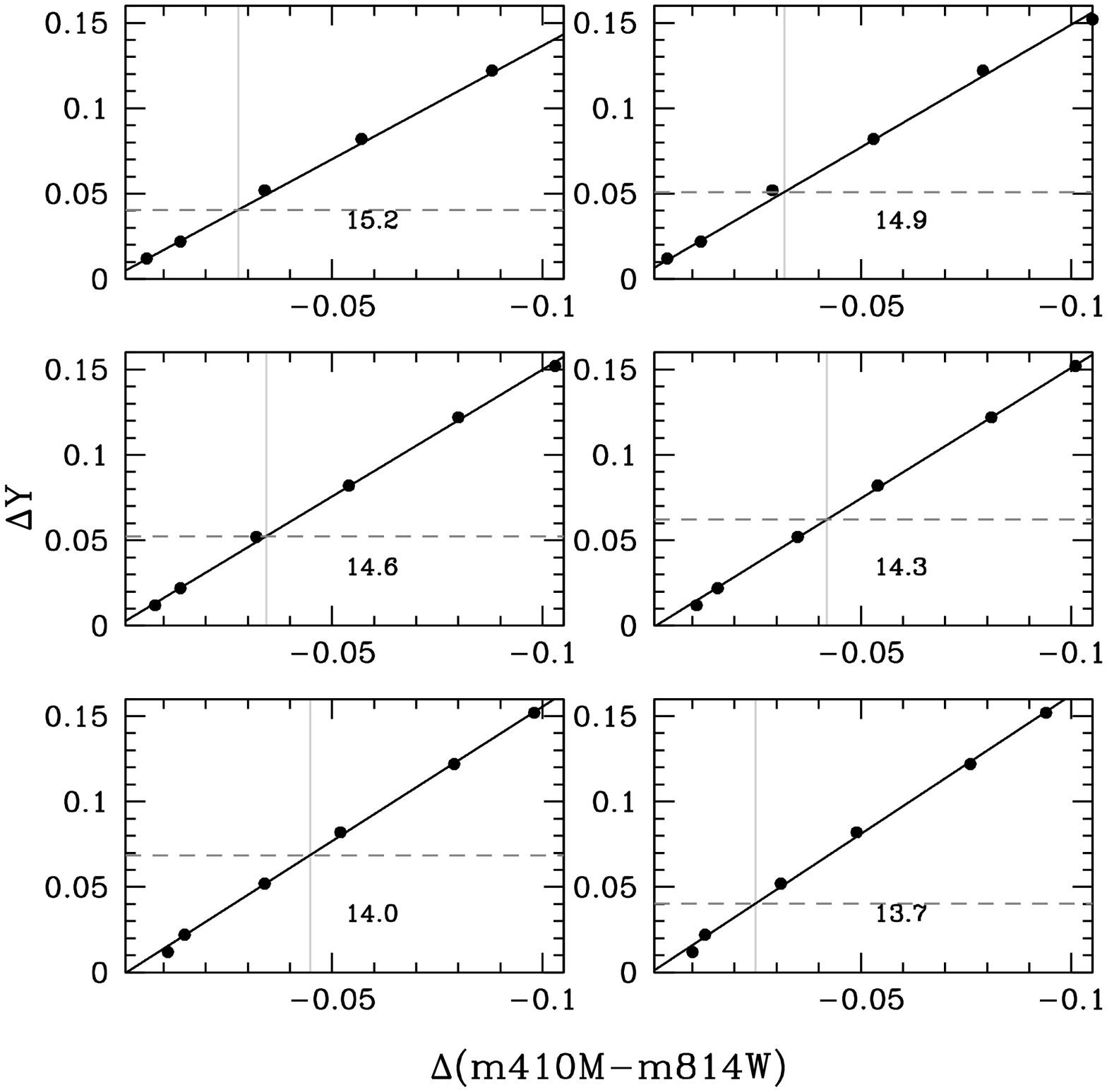}
\includegraphics[width=0.49\textwidth]{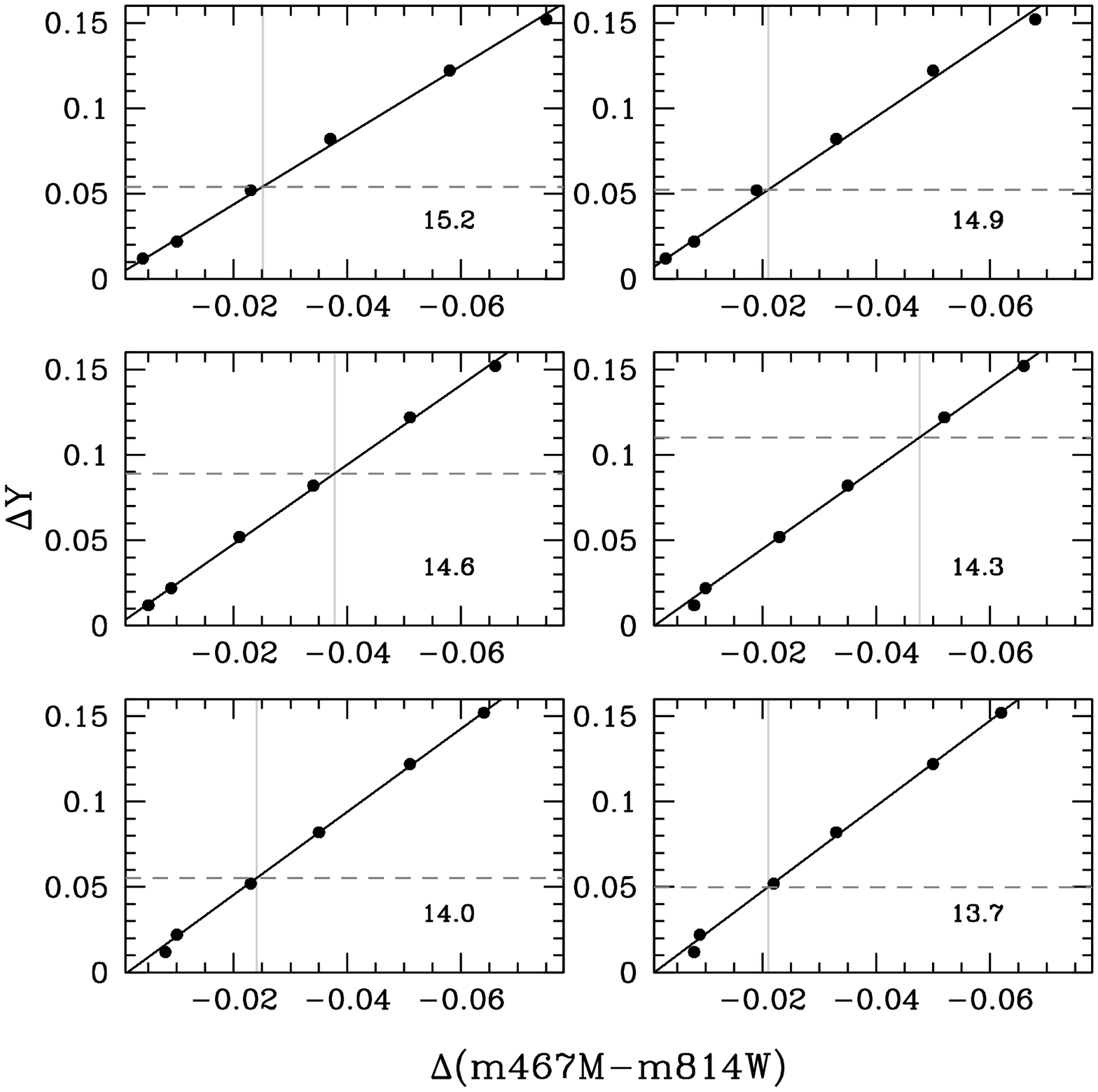}
\caption{Same as Fig.~\ref{395m814_1G2G_6752} (top panels) and Fig.~\ref{Ydcol395_6752} (bottom panels) for a the colors (m410M-m814W) (left panels) and (m467M-m814W) (right panels).}
\label{fig_deltaY}
\end{figure*}

\end{appendix}

\end{document}